 \documentstyle[12pt]{article}
 \newcommand{\nn}{\nonumber}
 \newcommand{\bd}{\begin{document}}
 \newcommand{\ed}{\end{document}}
 \newcommand{\bc}{\begin{center}}
 \newcommand{\ec}{\end{center}}
 \newcommand{\be}{\begin{eqnarray}}
 \newcommand{\ee}{\end{eqnarray}}
 \renewcommand{\thefootnote}{\alph{footnote}}
 \newcommand{\se}{\section}
 \newcommand{\sse}{\subsection}
 \newcommand{\bi}{\bibitem}
 \def\figcap{\section*{Figure Captions\markboth
      {FIGURECAPTIONS}{FIGURECAPTIONS}}\list
      {Figure \arabic{enumi}:\hfill}{\settowidth\labelwidth{Figure 999:}
      \leftmargin\labelwidth
      \advance\leftmargin\labelsep\usecounter{enumi}}}
 \let\endfigcap\endlist \relax
 \input psfig

\begin{document}
\baselineskip=0.75cm
 \begin{titlepage}

  \vskip 0.5in
  \null
 \begin{center}
  \vspace{.15in}
 {\LARGE {\bf Analysis of $B\rightarrow K^* \ell^+ \ell^-$ decays
  at large recoil }}\\ \vspace{1.0cm}
 \par
  \vskip 2.1em
  {\large
   \begin{tabular}[t]{c}
 {\bf Chuan-Hung Chen$^a$ and C.~Q.~Geng$^{b,c}$}
 \\
 \\
   {\sl${}^a$ Institute of Physics, Academia Sinica}
    \\ {\sl $\ $ Taipei, Taiwan 115, Republic of China }

 \\
 {\sl ${}^b$Department of Physics, National Tsing Hua University}
 \\  {\sl  $\ $ Hsinchu, Taiwan 300, Republic of China }
 \\
 {\sl ${}^c$ Theory Group, TRIUMF}
 \\ {\sl 4004 Wesbrrok Mall, Vancouver, B.C. V6T 2A3, Canada}
    \end{tabular}}
  \par \vskip 3.3em

 \date{\today}
  {\Large\bf Abstract}
 \end{center}
 We study the exclusive decays of $B\to K^{*}\ell^+ \ell^-$ within
 the framework of the perturbative QCD (PQCD). We obtain the form factors
for the $B\to K^{*}$ transition in the large recoil region, where the PQCD
 for heavy $B$ meson decays is reliable. We find that our results for
 the form factors at
 $q^2=0$ are consistent with those from most of the other QCD models in the
 literature.
 Via the decay chain of $B\rightarrow K^* (K\pi) \ell^+ \ell^-$, we obtain
 many physical observables related to the different
 helicity combinations of $B\rightarrow K^* \ell^+ \ell^-$.
 In particular, we point out that the T violating effect suppressed
 in the standard model can be up to $O(10\%)$ in some CP violating models
 with new physics.

 \end{titlepage}

 \section{Introduction}

 There has been an enormous progress for flavor physics
 since the CLEO observation \cite{cleo} of the radiative
 $b\to s\gamma $ decay.
  Recently, the decay modes of $B\to K \ell^+ \ell^-\ (\ell=e,\mu)$ have been
 observed \cite{Belle} at the Belle detector in the KEKB $e^+e^-$ storage
 ring with the branching ratio of
  $Br(B\to K \ell^+ \ell^-)=(0.75^{+0.25}_{-0.2
  1}\pm0.09)\times 10^{-6}$, while
 the standard model (SM) expectation is around $0.5\times 10^{-6}$
 \cite{Bdecays}. We remark that the decay has not yet been seen at
 the BaBar detector in the PEP-II B factory \cite{BaBar}.
 Experimental searches at the B-factories for $B\to K^* \ell^+
 \ell^-$ are also within the theoretical predicted ranges
 \cite{BCP43}. It is known that the study of flavor change neutral
 currents (FCNCs) in these B decays
  provides us with information on not only the
 Cabibbo-Kobayashi-Maskawa
 (CKM) quark-mixing matrix elements \cite{CKM} in the SM
 but also new physics such as supersymmetry (SUSY).

 On the other hand, via $B$ decays such as $B\to J/\Psi\,K$, we can
 test whether the unique phase in the CKM matrix is indeed the origin of
CP violation (CPV).
In general,
CP asymmetries (CPAs) in
 $B$ decays are defined by $a_{CP}\propto \Gamma
 -\bar{\Gamma}$ and $A_{CP}(t)\propto \Gamma(t)-\bar{\Gamma}(t)$,
called direct CPA or CP-odd observable and time dependent CPA,
respectively.
 The former
%, called direct CPA or CP-odd observable,
needs
 both weak CP violating and strong phases, while the latter
%one of the time dependent CPA
contains not only a non-zero CP-odd phase
 but also the $B-\bar{B}$ mixing. We note that the present world
 average for $a_{CP}^{\Psi K}$ is $0.79\pm0.12$  \cite{BCPV,SMBCPV}
 comparing with the SM prediction of $0.70\pm0.10$ \cite{SMBCPV}.
 In the decays of $B\to K^* \ell^+ \ell^-$, CPAs such as $a_{CP}$ are
 small even with  weak phases being $O(1)$ due to the smallness
of strong phases \cite{KL}.

 To study CPV in $B\to K^* \ell^+ \ell^-$, one can also define some
 T-odd observables by  momentum correlations,
% \cite{geng-PRL},
 such as the well known triple momentum correlations \cite{CGN}. These
 observables do not require strong phases in contrast to the CPA of
 $a_{CP}$.
 % and $A_{CP}(t)$.
 In the absence of final state
 interactions, these T-odd observables are T violating and thus CP
 violating by virtue of the CPT theorem. In the decays of $B\rightarrow
 K^{*} \ell^{+} \ell^{-}$ ($\ell=e,\mu$, and $\tau $), the spin $s$
 can be the polarized lepton, $s_{\ell},$ or the $K^{*} $ meson,
 $\epsilon ^{*}(\lambda )$. For the polarized lepton,
since the T-odd polarization is normally associated with the lepton mass,
  we expect that this type of T violating effects is
 suppressed and less than $1\%$ for the light lepton modes
 \cite{geng-PRD}. Although the $\tau$ mode can escape from the
 suppression, the corresponding branching ratio (BR) of
 $O(10^{-7})$ is about one order smaller than those of $e$ and
 $\mu$ modes. In this paper, we concentrate on the decay chain of
 $B\to K^* \ell^+ \ell^-\to K\pi \ell^+ \ell^-$ and give a
 systematic study on various possible physical observables,
 especially the T-odd ones.
 %we will show that these types of T violation are zero in the SM but they
 %could be sizable in new physics such as the theories with SUSY.

 It is known that one of the main theoretical uncertainties in
 studying exclusive hadron decays arises from the calculations of
 matrix elements. At the large momentum transfer $(q^2)$ region,
 Lepage and Brodsky (LB) \cite{LB1} have developed an approach
 based on the perturbative QCD (PQCD).
  In the LB formalism, the
 nonperturbative part is included in the hadron wave functions and
 the transition amplitude is factorized into the convolution of
 hadron wave functions and the hard amplitude of valence
 quarks. However, with the LB approach, it has been pointed out
 that the perturbative evaluation of the pion form factor suffers a
 non-perturbative enhancement in the end-point region with a
 momentum fraction $x\to 0$ \cite{IL}. If so, the hard amplitude is
 characterized by a low scale and the expansion in terms of
 a large coupling constant $ \alpha _{s}$ is not reliable.
 Furthermore, more serious end-point (logarithmic) singularities
 are observed in
 the $B\to \pi $ transition form factors \cite{SHB,ASY}
  from the twist-2 (leading-twist) contribution.
 The singularities become linear while including the twist-3
 (next-to-leading twist) wave function \cite{BF}. Because of these
 singularities, it was claimed that even at the low $q^{2}$ form factors are dominated by soft
 dynamics and not calculable in the PQCD \cite{KR}.

 Following the concept of the PQCD, if the spectator quark inside
 the $B$ meson with a momentum of $O(\bar{\Lambda})$, where
 $\bar{\Lambda}=M_{B}-m_{b}$ and $m_{b}$ is the $b$ quark mass,
 wants to catch up the outgoing quark with an energy of
 $O(M_{B}/2)$
 %%%%%%%%%%%%%%%%%
 %, which is the daughter of the $b$ quark decay,
 %%%%%%%%%%%%%%%%%
 to form a hadron, it should obtain a large energy from $b$ or
 the daughter of it. That is, hard gluons actually play an
 essential role in the $B$ meson with large energy released decays.
 Therefore, relevant decay amplitudes should be calculable
 perturbatively. It is clear that to deal with the problem of
singularities is the main part of the PQCD.
 In order to handle these singularities, the strategy of including
 $k_{T}$, the transverse momentum of the valence-quark \cite{LS},
 and threshold resummation \cite {S0,CT} have been proposed
 \cite{MPQCD}. It has been shown that the singularities do not
 exist in a self-consistent PQCD analysis \cite{MPQCD}. In the
 literature, the applications of this PQCD approach to the
 processes of $B\rightarrow PP$, such as $B\rightarrow K\pi $
 \cite{KLS1}, $ B\rightarrow \pi \pi $ \cite{pipi}, $B\rightarrow
 KK$ \cite{CL}, $B\rightarrow K\eta^{(\prime)}$ \cite{KS} and
 $B_{s}\rightarrow KK$ \cite{Chen1},
  as well as that of $B\rightarrow VP$, such as $B\rightarrow \phi
 \pi $ \cite{Melic}, $B\rightarrow \phi
 K$ \cite{CKLS}, $B\rightarrow \rho (\omega) \pi $ \cite{LY}
 and $B\rightarrow \rho (\omega) K $ \cite{Chen2}, have been
 studied and found that they are consistent with the experimental
 data. In this paper, to calculate the matrix elements of
 relevant current operators, we adopt the PQCD factorization
 formalism as
 \begin{eqnarray}
 \langle V|{\cal O}_{k}|B\rangle &=&\int [dx]\int \left[
 \frac{d^{2}\vec{b}}{ 4\pi }\right] \Phi _{V}^{*}(x_{2},\vec{b}
 _{2})\,  T_{k}(\{x\},\{\vec{b}\},M_{B})\Phi
 _{B}(x_{1},\vec{b}_{1}) \nonumber \\ &\times &S_{t}\left(
 \{x\}\right) \displaystyle{e^{-\displaystyle{S\left(
 \{x\},\{\vec{b}\},M_{B}\right) }}}\;  \label{eqpqcd}
 \end{eqnarray}
 where $\Phi _{V}^{*}\ (\Phi _{B})$ is the wave function of $V\ (B)$ meson,
 $T_{k}$ is the hard scattering amplitude dictated by
 relevant current
 operators, the exponential factor is the Sudakov factor \cite{CS,BS}, and $%
 S_{t}(x)$ \cite{Li1,KLS} expresses the threshold resummation
 factor.

 The paper is organized as follows. In Sec.~II, we study the form
 factors of the $B\to K^*$ transition in the framework of the PQCD.
 In Sec.~III, we write the angular distributions and define the
 physical observables for the decays of  $B\to K^*\ell^+
 \ell^-$. In Sec.~IV, we present the numerical analysis. We
 also compare our results
  with those in other QCD models. We give our
 conclusions in Sec.~V.

 \section{Form factors in $B\rightarrow K^{*}$}
 For decays with the $B\rightarrow K^*$ transition,  the $B$ meson momentum
 $p_{1}$ and the $K^*$ meson momentum $p_{2}$ and polarization vector
   $\epsilon$ in the $B$ meson rest frame and the light-cone coordinate are
 taken as
 \begin{eqnarray}
 p_{1}&=& \frac{M_{B}}{\sqrt{2}}(1,1,\vec{0}_{\bot}),\ \ \ p_{2}
 ={\frac{M_{B}}{\sqrt{2}\eta }}(\eta
 ^{2},r_{K^{*}}^{2},\vec{0}_{\bot }),  \nonumber \\
 \epsilon _{L} &=&{\frac{1}{\sqrt{2}r_{K^{*}}\eta }}(\eta ^{2},-r_{K^{*}}^{2},%
 \vec{0}_{\bot }), \ \ \ \epsilon _{T}(\pm )
 ={\frac{1}{\sqrt{2}}}(0,0,1,\pm i)
 \end{eqnarray}
 with $\eta \approx 1-q^{2}/M_{B}^{2}$ and
 $r_{K^*}=M_{K^*}/M_{B}$, while those for the spectators of $B$ and
 $K^*$  sides are expressed by
 \begin{eqnarray}
 k_{1}=\left(0, x_{1}{M_B \over \sqrt{2}}, \vec{k}_{1\bot}\right)
 \ \ {\rm and}\ \ k_{2}=\left(x_{2}{M_B \over \sqrt{2}}\eta, 0,
 \vec{k}_{2\bot}\right),
 \end{eqnarray}
 respectively. In our calculations, we will neglect the small
 contributions from  $m_{u,d,s}$ and
  $\bar{\Lambda}$ as well as $M^{2}_{K^*}$ due to the on-shell
 condition of the valence-quark preserved. From the
 results in Ref. \cite{BBKT}, the $K^*$ meson distribution
 amplitude up to twist-3 can be derived as follows:
 \begin{eqnarray}
 \langle K^*(p,\epsilon_{L})|\bar{s}(z)_{j}d(0)_{l}|0\rangle&=&
 {1\over \sqrt{2N_c}}\int^{1}_{0}dx e^{ixp\cdot
 z}\{M_{K^*}[\not{\epsilon}_{L}]_{lj}
 \phi_{K^*}(x)+[\not{\epsilon}_{L}\not{p}]_{lj}\phi^{t}_{K^*}(x)\nonumber
 \\
 && +M_{K^*}[I]_{lj}\phi^{s}_{K^*}(x)
 \}, \nonumber \\
 \langle K^*(p,\epsilon_{T})|\bar{s}(z)_{j}d(0)_{l}|0\rangle&=&
 {1\over \sqrt{2N_c}}\int^{1}_{0}dx e^{ixp\cdot
 z}\{M_{K^*}[\not{\epsilon}_{T}]_{lj}
 \phi_{K^*}^{v}(x)+[\not{\epsilon}_{T}\not{p}]_{lj}\phi^{T}_{K^*}(x)\nonumber
 \\&& +{M_{K^*}\over p\cdot n_{-}}i\varepsilon_{\mu\nu\rho\sigma}
 [\gamma_{5}\gamma^{\mu}]_{lj}\epsilon^{\nu}_{T}p^{\rho}n^{\sigma}_{-}
 \phi^{a}_{K^*}(x)
 \},\label{ss}
 \end{eqnarray}
 where $n_{-}=(0,1,\vec{0}_{\bot})$ and $\phi_{K^*}(x)$
 and $\phi^{T}_{K^*}(x)$ are the twist-2 wave functions for the
 longitudinal  and transverse components of the $K^*$ polarization,
 respectively, while
 the remaining wave functions belong to the twist-3 ones
with their explicit expressions given below.
 \subsection{Power counting}
 To show the $B\rightarrow K^*$ form factors, we first discuss
 the twist-3 contributions in the PQCD approach.
 As an illustration, we take
 %the result of Eq. (\ref{v}) as the example.
  %From Eq. (\ref{v}), we know that
 the integrand  of twist-2, the
 hard gluon exchange in the B meson side, to be
 \begin{eqnarray}
 I^{tw2}={\phi^{T}_{K^*}(x_{2})M^{2}_{B} \over
 [x_{1}x_{2}M^{2}_{B}+|\vec{k}_{2\bot}-\vec{k}_{1\bot}|^{2}]
 [x_{2}M^{2}_{B}+|\vec{k}_{2\bot}|^{2}]}
 \label{I2}
 \end{eqnarray}
 where the first term in  the denominator is the propagator of
 the exchanged hard gluon while the second one is that of
 the internal b-quark.
 As studied in Ref. \cite{MPQCD}, introducing
 $k_{\bot}$ degrees of freedom will bring large double logarithms
of  $\alpha_{s}\ln^{2}(k_{\bot}/M_{B})$ through radiative corrections.
 In order to improve the perturbative expansion, these effects should
 be resummed, called $k_{\bot}$ resummation \cite{CS,BS}.
 Consequently, the Sudakov form factor introduced will suppress
 the region of $k^{2}_{\bot}\sim O(\bar{\Lambda}^{2})$.
 According to the analysis of Ref. \cite{KLS}, via the Sudakov suppression,
  the average $<k^{2}_{2\bot}>$
 is of $O(\bar{\Lambda}M_{B})$ for $M_{B}\sim 5$ GeV. Hence,
 with including $k_{\bot}$ resummation effects, Eq. (\ref{I2}) becomes
 \begin{eqnarray}
 I^{tw2}={\phi^{T}_{K^*}(x_{2})M^{2}_{B} \over
 [x_{1}x_{2}M^{2}_{B}+O(\bar{\Lambda}M_{B})][x_{2}M^{2}_{B}+O(\bar{\Lambda}
 M_{B} )]}.
 \end{eqnarray}
 Since the fraction momentum $x_{1}$ is of
 $O(\bar{\Lambda}/M_{B})$ and
 %from Eq. (\ref{wavef}) we know that
 $\phi^{T}_{K^*}(x_{2})\propto x_{2}(1-x_{2})$,
it is easy to see that at
 the end-point $I^{tw2}$ behaves like
 \begin{eqnarray}
 I^{tw2}\sim {1\over \bar{\Lambda} M_{B}}.
 \end{eqnarray}
 On the other hand, the integrand of
 twist-3 is expressed as
 \begin{eqnarray}
 I^{tw3}={r_{K^*}\phi^{a}_{K^*}(x_{2})M^{2}_{B} \over
 [x_{1}x_{2}M^{2}_{B}+O(\bar{\Lambda}M_{B})][x_{2}M^{2}_{B}+O(\bar{\Lambda}
 M_{B} )]}\label{I3}\,.
 \end{eqnarray}
  From Ref. \cite{BBKT}, we find that the twist-3 wave
 function $\phi^{a}_{K^*}$ at the end-point is a constant so
 that
 %the behavior of $I^{tw3}$ becomes
 \begin{eqnarray}
 I^{tw3}&\longrightarrow & {r_{K^*}\over
 \bar{\Lambda}^{2}}=\frac{M_{K^*}}{\bar{\Lambda}}{1\over
 \bar{\Lambda} M_{B}}
 \end{eqnarray}
 as $x_{2}\rightarrow \bar{\Lambda}/M_{B}$. Hence,  the power
 behavior of $I^{tw3}$ in $M_{B}$ is the same as that of $I^{tw2}$.
 We note that since the twist-3 one contains  the most serious
 singularity
 (linear divergence), the contribution from a higher twist wave
 function, such as that of twist-4, should be the same as that of
 twist-3
 at most. However, by the definition of the twist wave function, we
 know that the twist-4 one is associated with a factor of $r^{2}_{K^*}$,
 and its contribution should be one power suppressed by $r_{K^*}$ than
 that of twist-3 so that it belongs to a higher power contribution in
 our consideration. In our analysis, its effect will be neglected.
 \subsection{Form Factors}

 We parametrize the $B\rightarrow K^{*}$ transition form factors
 with various types of interacting vertices as follows:
 \begin{eqnarray}
 \left\langle K^{*}(p_{2},\epsilon )\right| V_{\mu }\left| \bar{B}%
 (p_{1})\right\rangle &=&i\frac{V(q^{2})}{M_{B}+M_{K^{*}}}\varepsilon _{\mu
 \alpha \beta \rho }\epsilon ^{*\alpha }P^{\beta }q^{\rho },  \nonumber \\
 \left\langle K^{*}(p_{2},\epsilon )\right| A_{\mu }\left| \bar{B}%
 (p_{1})\right\rangle &=&2M_{K^{*}}A_{0}(q^{2})\frac{\epsilon ^{*}\cdot q}{%
 q^{2}}q_{\mu }+\left( M_{B}+M_{K^{*}}\right) A_{1}(q^{2})\left( \epsilon
 _{\mu }^{*}-\frac{\epsilon ^{*}\cdot q}{q^{2}}q_{\mu }\right)  \nonumber \\
 &&-A_{2}(q^{2})\frac{\epsilon ^{*}\cdot q}{M_{B}+M_{K^{*}}}\left( P_{\mu }-%
 \frac{P\cdot q}{q^{2}}q_{\mu }\right) ,  \nonumber \\
 \left\langle K^{*}(p_{2},\epsilon )\right| T_{\mu \nu }q^{\nu }\left| \bar{B}%
 (p_{1})\right\rangle &=&-iT_{1}(q^{2})\varepsilon _{\mu \alpha \beta \rho
 }\epsilon ^{*\alpha }P^{\beta }q^{\rho },  \nonumber \\
 \left\langle K^{*}(p_{2},\epsilon )\right| T_{\mu \nu }^{5}q^{\nu }\left|
 \bar{B}(p_{1})\right\rangle &=&T_{2}(q^{2})\left( \epsilon _{\mu }^{*}P\cdot
 q-\epsilon ^{*}\cdot qP_{\mu }\right) +T_{3}(q^{2})\epsilon ^{*}\cdot
 q\left( q_{\mu }-\frac{q^{2}}{P\cdot q}P_{\mu }\right)\,,
 \label{ff}
 \end{eqnarray}
 where $P=p_{1}+p_{2}$, $q=p_{1}-p_{2}$, $V_{\mu }=\ \bar{s}\
 \gamma _{\mu }\ b$, $A_{\mu }=\bar{s}\ \gamma _{\mu }\gamma _{5}\
 b$, $T_{\mu \nu }=\bar{s}\ i\sigma _{\mu \nu }b$, and $T_{\mu \nu
 }^{5}=\bar{s}\ i\sigma _{\mu \nu }\gamma _{5}\ b$. According to
 the PQCD factorization formalism shown in Eq. (\ref{eqpqcd}), the
 components of form factors defined in Eq. (\ref {ff}) up to
 twist-3 wave functions are given by
 \begin{eqnarray}
 V( q^{2}) &=&(1+r_{K^{*}})8\pi
 C_{F}M_{B}^{2}\int_{0}^{1}[dx]\int_{0}^{\infty
 }b_{1}db_{1}b_{2}db_{2}\ \phi
 _{B}( x_{1},b_{1})  \nonumber \\
 &&\times \Big\{ \Big[ \phi _{K^{*}}^{T}(x_{2})-r_{K^{*}}(
 x_{2}\phi _{K^{*}}^{v}(x_{2})-(\frac{2}{\eta }+x_{2})\phi
 _{K^{*}}^{a}(x_{2}))
 \Big]  \nonumber \\
 &&\times E( t_{e}^{( 1) }) h(
 x_{1},x_{2},b_{1},b_{2})  \nonumber \\
 && +r_{K^{*}}\Big[ \phi _{K^{*}}^{v}(x_{2})+\phi
 _{K^{*}}^{a}(x_{2})\Big] E( t_{e}^{( 2) }) h(
 x_{2},x_{1},b_{2},b_{1}) \Big\} ,  \label{v}
 \end{eqnarray}
 \begin{eqnarray}
 A_{0}( q^{2}) &=&8\pi
 C_{F}M_{B}^{2}\int_{0}^{1}[dx]\int_{0}^{\infty
 }b_{1}db_{1}b_{2}db_{2}\ \phi
 _{B}( x_{1},b_{1})  \nonumber \\
 &&\times \Big\{ \Big[ (1+\eta x_{2})\phi
 _{K^{*}}(x_{2})+r_{K^{*}}( (1-2x_{2})\phi
 _{K^{*}}^{t}(x_{2})\nonumber \\
  &&+(\frac{2}{\eta }-1-2x_{2})\phi
 _{K^{*}}^{s}(x_{2})) \Big] E( t_{e}^{( 1) }) h(
 x_{1},x_{2},b_{1},b_{2})  \nonumber \\
 && +2r_{K^{*}}\phi _{K^{*}}^{s}(x_{2})E( t_{e}^{( 2) }) h(
 x_{2},x_{1},b_{2},b_{1}) \Big\}\,, \label{a0}
 \end{eqnarray}
 \begin{eqnarray}
 A_{1}( q^{2}) &=&\frac{8\pi C_{F}M_{B}^{2}\eta }{1+r_{K^{*}}}%
 \int_{0}^{1}[dx]\int_{0}^{\infty }b_{1}db_{1}b_{2}db_{2}\ \phi
 _{B}t(
 x_{1},b_{1})  \nonumber \\
 &&\times \Big\{ \Big[ \phi _{K^{*}}^{T}(x_{2})+r_{K^{*}}( (
 \frac{2}{\eta }+x_{2}) \phi _{K^{*}}^{v}(x_{2})-x_{2}\phi
 _{K^{*}}^{a}(x_{2})) \Big]  \nonumber \\
 &&\times E( t_{e}^{( 1) }) h(
 x_{1},x_{2},b_{1},b_{2})  \nonumber \\
 && +r_{K^{*}}\Big[ \phi _{K^{*}}^{v}(x_{2})+\phi
 _{K^{*}}^{a}(x_{2})\Big] E( t_{e}^{( 2) }) h(
 x_{2},x_{1},b_{2},b_{1}) \Big\} ,  \label{a1}
 \end{eqnarray}
 \begin{eqnarray}
 A_{2}( q^{2}) &=&\frac{(1+r_{K^{*}})^{2}}{\eta }A_{1}(
 q^{2}) -\frac{1+r_{K^{*}}}{\eta }2r_{K^{*}}A_{0}(q^{2})  \nonumber \\
 &&-\frac{1+r_{K^{*}}}{\eta }32r_{K^{*}}\pi
 C_{F}M_{B}^{2}\int_{0}^{1}[dx]\int_{0}^{\infty }b_{1}db_{1}b_{2}db_{2}\ \phi
 _{B}( x_{1},b_{1})  \nonumber \\
 &&\times ( (\frac{1}{\eta }-1)\phi _{K^{*}}^{t}(x_{2})-(\frac{1}{\eta }%
 -1)\phi _{K^{*}}^{s}(x_{2})) E( t_{e}^{( 1) }) h(
 x_{1},x_{2},b_{1},b_{2})\,,
   \label{a2}
 \end{eqnarray}
 \begin{eqnarray}
 T_{1}( q^{2}) &=&8\pi
 C_{F}M_{B}^{2}\int_{0}^{1}[dx]\int_{0}^{\infty
 }b_{1}db_{1}b_{2}db_{2}\ \phi
 _{B}( x_{1},b_{1})  \nonumber \\
 &&\times \Big\{ \Big[ ( 1+\eta x_{2}) \phi
 _{K^{*}}^{T}(x_{2})+r_{K^{*}}( (1-2x_{2})\phi
 _{K^{*}}^{v}(x_{2})  \nonumber \\
 &&+(\frac{2}{\eta }-1-2x_{2})\phi _{K^{*}}^{a}(x_{2})) \Big] E(
 t_{e}^{( 1) }) h(
 x_{1},x_{2},b_{1},b_{2})  \nonumber \\
 && +r_{K^{*}}\Big[ \phi _{K^{*}}^{v}(x_{2})+\eta \phi
 _{K^{*}}^{a}(x_{2})\Big] E( t_{e}^{( 2) }) h(
 x_{2},x_{1},b_{2},b_{1}) \Big\} ,  \label{t1}
 \end{eqnarray}
 \begin{eqnarray}
 T_{2}( q^{2}) &=&8\pi
 C_{F}M_{B}^{2}\int_{0}^{1}[dx]\int_{0}^{\infty
 }b_{1}db_{1}b_{2}db_{2}\phi
 _{B}( x_{1},b_{1})  \nonumber \\
 &&\times \Big\{ \Big[ (1+\eta x_{2})\phi
 _{K^{*}}^{T}(x_{2})+r_{K^{*}}( 2-\eta (1+2x_{2})) \phi
 _{K^{*}}^{v}(x_{2})\nonumber \\
 && +r_{K^{*}}\eta ( 1-2x_{2}) \phi _{K^{*}}^{a}(x_{2})\Big] E(
 t_{e}^{( 1) }) h(
 x_{1},x_{2},b_{1},b_{2})  \nonumber \\
 &&+r_{K^{*}}\eta \Big[ \phi _{K^{*}}^{v}(x_{2})+\phi
 _{K^{*}}^{a}(x_{2})\Big] E( t_{e}^{( 2) }) h(
 x_{2},x_{1},b_{2},b_{1}) \Big\} ,  \label{t2}
 \end{eqnarray}
 \begin{eqnarray}
 T_{3}( q^{2}) &=&8\pi
 C_{F}M_{B}^{2}\int_{0}^{1}dx_{1}dx_{2}\int_{0}^{\infty
 }b_{1}db_{1}b_{2}db_{2}\ \phi _{B}( x_{1},b_{1})  \nonumber \\
 &&\times \Big\{ \Big[ (1+\eta x_{2})\phi
 _{K^{*}}^{T}(x_{2})+r_{K^{*}}\Big( \frac{2}{\eta }-( 1+2x_{2}
 ) \Big) \phi _{K^{*}}^{v}(x_{2}) \nonumber \\
 &&+r_{K^{*}}( (1-2x_{2})) \phi _{K^{*}}^{a}(x_{2})-\frac{
 2r_{K^{*}}}{\eta }( \phi _{K^{*}}(x_{2})\nonumber \\
 &&+r_{K^{*}}( \frac{2}{\eta }+x_{2}) \phi
 _{K^{*}}^{t}(x_{2})-r_{K^{*}}x_{2}\phi _{K^{*}}^{s}(x_{2})
 )
 \Big] E( t_{e}^{( 1) }) h(
 x_{1},x_{2},b_{1},b_{2})  \nonumber \\
 &&+r_{K^{*}}\Big[ \phi _{K^{*}}^{v}(x_{2})+\phi _{K^{*}}^{a}(x_{2})-%
 \frac{2r_{K^{*}}}{\eta }\phi _{K^{*}}^{s}(x_{2})\Big] E( t_{e}^{(
 2) }) h( x_{2},x_{1},b_{2},b_{1}) \Big\} . \label{t3}
 \end{eqnarray}
 The evolution factor is given by
 \begin{equation}
 E\left( t\right) =\alpha _{s}\left( t\right) \exp \left( -S_{B}\left(
 t\right) -S_{K^{*}}\left( t\right) \right) \,,  \label{ef}
 \end{equation}
 where explicit expressions of the Sudakov exponents $S_{B(K^{*})}$
 can be found in Ref. \cite{CL}. The hard function of  $h$ is written as
 \begin{eqnarray}
 h(x_{1},x_{2},b_{1},b_{2}) &=&S_{t}(x_{2})K_{0}\left( \sqrt{x_{1}x_{2}\eta }%
 M_{B}b_{1}\right)  \nonumber \\
 &&\times \left[ \theta (b_{1}-b_{2})K_{0}\left( \sqrt{x_{2}\eta }%
 M_{B}b_{1}\right) I_{0}\left( \sqrt{x_{2}\eta }M_{B}b_{3}\right)
 \right. \nonumber \\ &&\left. +\theta (b_{2}-b_{1})K_{0}\left(
 \sqrt{x_{2}\eta }M_{B}b_{2}\right) I_{0}\left( \sqrt{x_{2}\eta
 }M_{B}b_{1}\right) \right]   \label{dh}
 \end{eqnarray}
 where the threshold resummation effect is described by \cite{KLS}
 \begin{eqnarray*}
 S_{t}(x)={2^{1+2c}\Gamma(\frac{3}{2}+c) \over
 \sqrt{\pi}\Gamma(1+c)}[x(1-x)]^{c}.
 \end{eqnarray*}
  The hard scales $t^{(1,2)}$ are chosen to be
 \begin{eqnarray*}
 t^{( 1) } &=&\max ( \sqrt{M_{B}^{2}\eta x_{2}}
 ,1/b_{1},1/b_{2}) \,,  \nonumber \\
 t^{( 2) } &=&\max ( \sqrt{M_{B}^{2}\eta x_{1}}%
 ,1/b_{1},1/b_{2}) \,.  \label{tscale}
 \end{eqnarray*}
 For the $K^{*}$ meson distribution amplitudes, we adopt the results given
 in Ref. \cite{BBKT} and explicitly we have
 \begin{eqnarray}
 \phi _{K^{*}}(x) &=&\frac{3f_{K^{*}}}{\sqrt{2N_{c}}}x( 1-x) [
 1+0.57( 1-2x) +0.1( 5( 1-2x) ^{2}-1) ] , \nonumber \\
 \phi _{K^{*}}^{t}(x) &=&\frac{f_{K^{*}}^{T}}{2\sqrt{2N_{c}}}[ 0.3(
 1-2x) ( 3( 1-2x) ^{2}+10( 1-2x) -1)  \nonumber\\
 &&+0.06( 1-2x) ^{2}( 5( 1-2x) ^{2}-3) +0.21( 3-30(
 1-2x) ^{2}+35( 1-2x) ^{4})\nonumber \\
 && +0.36( 1-2( 1-2x) ( 1+\ln (1-x)) ) ] , \nonumber \\
 \phi _{K^{*}}^{s}(x) &=&\frac{3f_{K^{*}}^{T}}{2\sqrt{2N_{c}}}[ (
 1-2x) ( 1+0.2( 1-2x) +0.6( 10x^{2}-10x+1) ) \nonumber\\
 &&
 -0.4x(1-x)+0.12( 1-6x-2\ln (1-x)) ] ,\nonumber \\
 \phi _{K^{*}}^{T}(x) &=&\frac{3f_{K^{*}}^{T}}{\sqrt{2N_{c}}}x(
 1-x) [ 1+0.60( 1-2x) +0.06( 5( 1-2x)
 ^{2}-1) ] ,\nonumber \\
 \phi _{K^{*}}^{v}(x) &=&\frac{f_{K^{*}}}{2\sqrt{2N_{c}}}[ \frac{3}{4}%
 ( 1+( 1-2x) ^{2}+0.44( 1-2x) ^{3})
 +0.20( 3( 1-2x) ^{2}-1)  \nonumber\\
 && +0.11( 3-30( 1-2x) ^{2}+35( 1-2x) ^{4}) +0.48( 2x+\ln (1-x)) ]
 ,\nonumber
 \\ \phi _{K^{*}}^{a}(x) &=&\frac{3f_{K^{*}}}{4\sqrt{2N_{c}}}[
 ( 1-2x) ( 1+0.19( 1-2x) +0.81( 10x^{2}-10x+1) ) \nonumber
 \\ && - 1.14x(1-x)+0.16( 1-6x-2\ln (1-x)) ]
 \,. \label{wavef}
 \end{eqnarray}
  From Eqs. (\ref{a2})-(\ref{t2}),
 at $q^{2}=0$ we obtain the identities
 \begin{eqnarray}
 A_{2}(0) &=&( 1+r_{K^{*}}) ^{2}A_{1}(0)-2r_{K^{*}}( 1+r_{K^{*}})
 A_{0}(0), \label{idff}
 \nn\\
 T_{1}(0) &=&T_{2}(0),
 \end{eqnarray}
 which are consistent with
the leading order model-independent relation
\cite{BF,Charles,Burdman,Ali,MS,LF}
 \begin{eqnarray}
 A_{2}(0) &=&{1+r_{K^{*}}\over 1-r_{K^*}} A_{1}(0)-{2r_{K^{*}}
 \over 1-r_{K^*}}  A_{0}(0)\,.
 \end{eqnarray}
%  from other QCD models in Refs. \cite{Ali,MS,LF}.
  We note that due to the parametrization of Eq.
 (\ref{ff}), there are terms proportional to
 $r^{2}_{K^*}$  in Eqs.
 (\ref{v}), (\ref{a1}) and (\ref{a2}). In order to guarantee that
 only $r_{K^*}$ dependence appears in the left-handed sides of Eq.
 (\ref{ff}), those with $r^{2}_{K^*}$ should not be dropped.
 \section{Angular distributions and physical observables}

 \subsection{Effective Hamiltonians and Decay Amplitudes}

 The effective Hamiltonians of $b\rightarrow s\ \ell^{+}\ell^{-}$
 are given by \cite{Buras}

 \begin{equation}
 {\cal H}=\frac{G_{F}\alpha \lambda _{t}}{\sqrt{2}\pi }\left[ H_{1\mu }L^{\mu
 }+H_{2\mu }L^{5\mu }\right]  \label{heff}
 \end{equation}
 with
 \begin{eqnarray}
 H_{1\mu } &=&C_{9}(\mu )\bar{s}\gamma _{\mu }(\mu )P_{L}b\ -\frac{2m_{b}}{%
 q^{2}}C_{7}(\mu )\bar{s}i\sigma _{\mu \nu }q^{\nu }P_{R}b \,,
 \nn\\
 H_{2\mu } &=&C_{10}\bar{s}\gamma _{\mu }P_{L}b \,,
 \nn\\
 L^{\mu } &=&\bar{\ell}\gamma ^{\mu }\ell\,, \nn\\
 L^{5\mu } &=&\bar{\ell}\gamma ^{\mu }\gamma _{5}\ell\,,
 \end{eqnarray}
 where $\lambda_t=V_{tb}V_{ts}^*$ and $C_{9}(\mu )$, $C_{10}$ and
 $C_{7}(\mu )$ are the Wilson coefficients (WCs)
and their expressions can
be found in
 Ref. \cite{Buras} for the SM. Since the operator associated with
 $C_{10}$ is not renormalized under QCD, it is the only one with
 the $\mu $ scale free. Besides the short-distance (SD)
 contributions, the main effect on the branching ratios comes from
 c\={c} resonant states such as $\Psi ,\Psi ^{\prime }\,,
  etc.,$ $i.e.$, the long-distance (LD) contributions.
 In the literature \cite{DTP,LMS,AMM,OT,Sehgal}, it has been
 suggested by combining the FA and the vector meson dominance (VMD)
 approximation to estimate LD effects for the $B$ decays. With
 including the resonant effect (RE) and absorbing it to the related
 WC, we obtain the effective WC of $C_{9}$ as
 \begin{equation}
 C_{9}^{eff}=C_{9}\left( \mu \right) +\left( 3C_{1}\left( \mu \right)
 +C_{2}\left( \mu \right) \right) \left( h\left( x,s\right) +
\frac{3}{\alpha^2 }%
 \sum_{j=\Psi ,\Psi ^{\prime }}k_{j}\frac{\pi \Gamma \left( j\rightarrow
 l^{+}l^{-}\right) M_{j}}{q^{2}-M_{j}^{2}+iM_{j}\Gamma _{j}}\right) \,,
 \label{effc9}
 \end{equation}
 where
 $h(x,s)$ describes
 the one-loop matrix elements of operators $O_{1}=\bar{s}_{\alpha }\gamma
 ^{\mu }P_{L}b_{\beta }\ \bar{c}_{\beta }\gamma _{\mu }P_{L}c_{\alpha }$ and $%
 O_{2}=\bar{s}\gamma ^{\mu }P_{L}b\ \bar{c}\gamma _{\mu }P_{L}c$ \cite{Buras}%
 , $M_{j}$ ($\Gamma _{j}$) are the masses (widths) of intermediate states,
 and the factors $k_{j}$ are phenomenological parameters for compensating the
 approximations of FA and VMD and reproducing the correct branching ratios $%
 Br\left( B\rightarrow J/\Psi X\rightarrow l^{+}l^{-}X\right) =Br\left(
 B\rightarrow J/\Psi X\right) $ $\times $ $Br\left( J/\Psi \rightarrow
 l^{+}l^{-}\right) $. For simplicity, we neglect the small WCs
 and take $k_{j}=-1/\left( 3C_{1}\left(
 \mu \right) +C_{2}\left( \mu \right) \right)$.
It is clear
 that the uncertainty related to this assumption can be large \cite{LW}.
Moreover, it is questionable whether one can include both
quark-level calculations with $c\bar{c}$-loop and resonances in
Eq. (\ref{effc9}). However, since we are only interested in
physics behind the various observables at the large recoil we
shall not discuss the uncertainties arising from Eq.
(\ref{effc9}).

%\cite{LW}
%Z.~Ligeti and M.~B.~Wise,
%%``$$|$V_{ub}$|$$ from exclusive $B$ and $D$ decays,''
%Phys.\ Rev.\ D {\bf 53}, 4937 (1996)

 Combining Eqs. (\ref{ff}) and (\ref{heff}), the transition amplitudes for
 $%
 B\rightarrow K^{*}\ell^{+} \ell^{-}$ ($\ell=\mu ,\ e$) can be
 written as
 \begin{equation}
 {\cal M}_{K^{*}}^{(\lambda )}=\frac{G_{F}\alpha \lambda _{t}}{2\sqrt{2}%
 \pi }\left\{ {\cal M}_{1\mu }^{(\lambda )}L^{\mu }+{\cal M}_{2\mu
 }^{(\lambda )}L^{5\mu }\right\}  \label{ampk*}
 \end{equation}
 with
 \begin{eqnarray}
 {\cal M}_{1\mu }^{(\lambda )} &=&ih_{1}\varepsilon _{\mu \nu
 \alpha \beta }\epsilon ^{*\nu }(\lambda )P^{\alpha }q^{\beta
 }+h_{2}\epsilon _{\mu }^{*}(\lambda )+h_{3}\epsilon ^{*}\cdot
 qP_{\mu },  \nonumber \\ {\cal M}_{2\mu }^{(\lambda )}
 &=&ig_{1}\varepsilon _{\mu \nu \alpha \beta }\epsilon ^{*\nu
 }(\lambda )P^{\alpha }q^{\beta }+g_{2}\epsilon _{\mu }^{*}(\lambda
 )+g_{3}\epsilon ^{*}\cdot qP_{\mu },
 \end{eqnarray}
 \begin{eqnarray} h_{1} &=&{V^{9}(q^{2})\over m_{B}+m_{K^*}}
 +\frac{2m_{b}}{q^{2}}(\mu
 )T^{7}_{1}(q^{2}), \nonumber \\
 h_{2} &=&-(m_{B}+m_{K^*})A^{9}_{1}(q^{2})-\frac{2m_{b}}{q^{2}}
 P\cdot q T^{7}_{2} (q^{2}),  \nonumber \\
 h_{3} &=&{A^{9}_{2}(q^{2})\over
 m_{B}+m_{K^*}}+\frac{2m_{b}}{q^{2}}
 \Big(T^{7}_{2}(q^{2})+\frac{q^2}{P\cdot q}T^{7}_{3}(q^2)\Big),
 \nonumber \\ g_{1} &=&C_{10}{V(q^{2})\over m_{B}+m_{K^*}}, \nonumber \\
 g_{2} &=&-C_{10}(m_{B}+m_{K^*})A_{1}(q^{2})\,,
  \nonumber \\
 g_{3} &=&C_{10}{A_{2}(q^{2}) \over m_{B}+m_{K^*}}\,,
  \label{ampk1}
 \end{eqnarray}
 where the form factors associated with superscripts denote the relevant
 WCs convoluted with hard amplitudes and wave
 functions, described by
 \begin{eqnarray}
 F^{k}(q^2)=\int[dx][db]\Phi^{*}_{K^*}(x_{2},\vec{b}_{2}) C_{k}(t)
 T_{\mu}(\{x\},\{\vec{b}\},M_{B})\Phi_{B}(x_{1},\vec{b}_{1})S_{t}
 (\{x\})e^{-S(\{x\},\{\vec{b}\},M_{B})}.
 \label{wff}
 \end{eqnarray}
  It is worth to mention that for the convenient in the PQCD
 formalism, Eq. (\ref{wff}) can be written as $F^{k}\approx
 C_{k}(t_0)\,F$ with F being the form factor and
$t_{0}=\sqrt{\bar{\Lambda}M_{B}}$.
 \subsection{Angular Distributions}

 In the literature, there are a lot of discussions on $B\rightarrow
 K^* \ell^{+} \ell^{-}$ decays. However, most of them have
been concentrated
 on the differential decay rates and lepton
 %longitudinal and transverse
 polarization and forward-backward asymmetries.
  It is known that the differential decay rates have large
 uncertainties from not only hadronic matrix elements but also the
 parametrizations of LD effects, and the lepton polarization asymmetries
 are hard to be observed due to the difficulties of
  measuring lepton polarizations.
 Therefore,  to test the SM and search for new physics in
 $B\rightarrow K^* \ell^{+} \ell^{-}$, it is necessary to find some
 other physical observables which  have less theoretical
uncertainties but measurable experimentally, similar to the
zero positions in the forward-backward asymmetries \cite{Ali,Bu}.
 It is found that if one considers the decay chain of $%
 B\rightarrow K^{*} \ell^{+} \ell^{-}\rightarrow (K\pi )\ell^{+}
 \ell^{-}$, via the study of different angular distributing
 components, we can analyze (a) contributions from both
 longitudinal and transverse parts of the $K^*$ polarization, (b)
 T-even (CP conserved)  effects from the mixings of
 longitudinal and transverse polarizations of $K^*$, and (c) T-odd
  effects from the mixings in (b). We note that some T-odd effects are
 suppressed in the SM and thus, measuring these effects could
 indicate CP violation from new physics \cite{TVCQ}.

 To understand dynamical dependence in T-odd terms of
 $\varepsilon_{\mu\nu\alpha\beta}
 q^{\mu}\epsilon^{*\nu}(\lambda)p_{\ell}^{\alpha}P^{\beta}$, it is
 inevitable to investigate the processes of $B\rightarrow K^{*} \ell^{+}
 \ell^{-}\rightarrow (K\pi )\ell^{+}\ell^{-}$ so that the
 polarization $\lambda $ and $\lambda ^{\prime }$ in the
 differential decay rates, written as $d\Gamma \propto H(\lambda
 ,\lambda ^{\prime })$ ${\cal M}_{K^{*}}^{(\lambda )}$ ${\cal
 M}_{K^{*}}^{(\lambda ^{\prime })\dagger }$ with $H(\lambda
 ,\lambda ^{\prime })\equiv \epsilon (\lambda )\cdot p_{K}$
 $\epsilon ^{*}(\lambda ^{\prime })\cdot p_{K}$, can be different.
  From Eq. (\ref{ampk*}), we see that ${\cal M}_{2\mu }^{(\lambda )}
 $ only depends on $C_{10}$. Clearly, T violating effects can
 not be generated from ${\cal M}_{2\mu }^{(\lambda )} {\cal
 M}_{2\mu ^{\prime }}^{(\lambda ^{\prime })\dagger}$, but induced
  from ${\cal M}_{1\mu }^{(\lambda )} {\cal M}_{1\mu^{\prime
 }}^{(\lambda ^{\prime })\dagger }$ as well as  ${\cal M}_{1\mu}^{(\lambda
 )} {\cal M}_{2\mu ^{\prime }}^{(\lambda ^{\prime
 })\dagger }$. This can be understood as follows: firstly, for the ${\cal M}%
 _{1\mu }^{(\lambda )}{\cal M}_{1\mu ^{\prime }}^{(\lambda ^{\prime
 })\dagger }TrL^{\mu }L^{\mu ^{\prime }}$ part with
 $TrL^{\mu }L^{\mu ^{\prime }}\sim ( p_{\ell^{-}}^{\mu
 }p_{\ell^{+}}^{\mu ^{\prime }}+p_{\ell^{-}}^{\mu ^{\prime
 }}p_{\ell^{+}}^{\mu }-g^{\mu \mu ^{\prime }}p_{\ell^{-}}\cdot
 p_{\ell^{+}}) $, the relevant T-odd terms can be roughly expressed
 by
 \begin{eqnarray}
 {\cal M}_{1\mu }^{(\lambda )}{\cal M}_{1\mu ^{\prime }}^{(\lambda
 ^{\prime})\dagger }TrL^{\mu }L^{\mu ^{\prime }}& \propto&
 Z_{1}{Im} h_{1}h_{3}^{*}\epsilon (0)\cdot q\varepsilon _{\mu \nu
 \alpha \beta }q^{\mu }\epsilon ^{*\nu }(\pm )p_{\ell^{+}}^{\alpha
 }P^{\beta } \nonumber \\ &&+Z_{2}{Im}h_{1}h_{2}^{*}\epsilon
 (0)\cdot p_{\ell^{+}}\varepsilon _{\mu \nu \alpha \beta }q^{\mu
 }\epsilon ^{*\nu }(\pm )p_{\ell^{+}}^{\alpha }P^{\beta } \nonumber
 \\ &&+Z_{3}{Im}h_{1}h_{2}^{*}\epsilon (\mp )\cdot
 p_{\ell^{+}}\varepsilon _{\mu \nu \alpha \beta }q^{\mu }\epsilon
 ^{*\nu }(\pm )p_{\ell^{+}}^{\alpha }P^{\beta } \label{im1}
 \end{eqnarray}
 where $Z_i\ (i=1,2,3)$ are functions of kinematic variables.
  From Eq. (\ref{ampk1}), one gets ${Im}%
 h_{1}h_{2}^{*}\sim{Im} h_{1}h_{3}^{*}\sim {Im}C_{9}^{eff} (\mu
 )C_{7}(\mu )$ . We note that, as shown in Eq. (\ref{im1}), the
 T-odd observables could be non-zero if the processes involve strong
 phases or absorptive parts even without CP violating phases. By
 means of Eq. (\ref{effc9}), $C_{9}^{eff}(\mu )$ includes the
 absorptive parts such that the results of Eq. (\ref{im1}) do not
 vanish in the SM. Secondly, for ${\cal M}_{1\mu}^{(\lambda )}
 {\cal M}_{2\mu ^{\prime }}^{(\lambda ^{\prime })\dagger }TrL^{\mu
 }L^{5\mu ^{\prime }}$, one gets
 \begin{eqnarray}
 ( {\cal M}_{1\mu }^{(\lambda )}{\cal M}_{2\mu ^{\prime
 }}^{(\lambda ^{\prime })\dagger }+{\cal M}_{2\mu }^{(\lambda
 )}{\cal M}_{1\mu ^{\prime }}^{(\lambda ^{\prime })\dagger })
 TrL^{\mu }L^{5\mu ^{\prime }} \propto   (
 {Im}h_{2}g_{3}^{*}-{Im}h_{3}g_{2}^{*}) \varepsilon _{\mu \nu
 \alpha \beta }q^{\mu }\epsilon ^{*\nu }(\pm )p_{\ell^{+}}^{\alpha
 }P^{\beta } \label{im2}
 \end{eqnarray}
 where $TrL^{\mu }L^{5\mu ^{\prime }}=-4i\varepsilon ^{\mu \mu
 ^{\prime }\alpha \beta }q_{\alpha }p_{l^{+}\beta }$ has been used.
  From Eq. (\ref {ampk1}), we find that
 ${Im}h_{2}g_{3}^{*}-{Im}h_{3}g_{2}^{*}$ is only related to
 ${Im}C_{7}(\mu )C_{10}^{*}$ and the dependence of ${Im}C_{9}(\mu
 )C_{10}^{*}$ is canceled in Eq. (\ref{im2}). For the decays of $
 b\rightarrow s\ell^{+}\ell^{-}$, since  the absorptive parts in $
 C_{7}(\mu)$ and $C_{10}$ are not expected, a non-vanishing value
 of ${Im}C_{7}(\mu )C_{10}^{*} $ indicates pure weak CP
 violating effects.

  In order to derive the whole differential decay rates with the
 $K^{*}$ polarization, we choose that  $K^*$ helicities are
 $\epsilon (0)=(|\vec{p}_{K^{*}}|,0,0,E_{K^{*}})/M_{K^{*}}$ and
 $\epsilon (\pm )=(0,1,\pm i,0)/\sqrt{2}$, the positron lepton
 momentum $p_{\ell^{+}}=\sqrt{q^{2}}(1,\sin \theta _{\ell},0,\cos
 \theta _{\ell})/2$ with
 $E_{K^{*}}=(M_{B}^{2}-M_{K^{*}}^{2}-q^{2})/2\sqrt{q^{2}}$ and
 $|\vec{p}_{K^{*}}|=\sqrt{E_{K^{*}}^{2}-M_{K^{*}}^{2}}$ in the
 $q^{2}$ rest frame,  and the $K$ meson momentum  $p_{K}=(1,\sin
 \theta _{K}\cos \phi ,\sin \theta _{K}\sin \phi ,\cos \theta
 _{K})M_{K^{*}}/2$ in the $K^{*}$ rest frame where $\phi $ denotes
 the relative angle of the decaying plane between $K\pi $ and
 $\ell^{+}\ell^{-}$. From Eq. (\ref{ampk*}), the
 differential decay rates of $B\rightarrow
 K^{*}\ell^{+}\ell^{-}\rightarrow (K\pi )\ell^{+}\ell^{-}$
  as functions of angles $\theta _{\ell}$,
 $\theta _{K} $ and $\phi $ are given by
 \begin{eqnarray}
 \frac{d\Gamma }{d\cos \theta _{K}d\cos \theta _{\ell}d\phi dq^{2}}
 %\nn\\
 &=&\frac{ 3\alpha _{em}^{2}G_{F}^{2}| \lambda _{t}| ^{2}| \vec{p}
 | }{2^{14}\pi ^{6}M_{B}^{2}}Br(K^{*}\rightarrow K\pi ) \nonumber
 \\ &&\times \Big\{ 4\cos ^{2}\theta _{K}\sin ^{2}\theta
 _{\ell}\sum_{i=1,2}| {\cal M}_{i}^{0}| ^{2} \nonumber \\
 && +\sin ^{2}\theta _{K}(1+\cos ^{2}\theta _{\ell})\sum_{i=1,2}( |
 {\cal M}_{i}^{+}|
 ^{2}+| {\cal M}_{i}^{-}| ^{2}) \nonumber \\
 &&-\sin 2\theta _{K}\sin 2\theta _{\ell}\Big[ \cos \phi
 \sum_{i=1,2}{Re}( {\cal M}_{i}^{+}+{\cal M}_{i}^{-}) {\cal
 M}_{i}^{0*} \nonumber \\
 && +\sin \phi \sum_{i=1,2}{Im}( {\cal M}_{i}^{+}-{\cal M}
 _{i}^{-}) {\cal M}_{i}^{0*}\Big]  \nonumber \\
 &&-2\sin ^{2}\theta _{K}\sin ^{2}\theta _{\ell}\Big[ \cos 2\phi
 \sum_{i=1,2}{Re }( {\cal M}_{i}^{+}{\cal M}_{i}^{-*}) +\sin 2\phi
 \sum_{i=1,2}{Im}
 ( {\cal M}_{i}^{+}{\cal M}_{i}^{-*}) \Big]  \nonumber \\
 &&-2\sin ^{2}\theta _{K}\cos \theta _{\ell}\Big[ 2{Re}{\cal
 M}_{1}^{+}{\cal M}_{2}^{+*}-2{Re}{\cal M}_{1}^{-}{\cal M}_{2}^{-*}\Big]  \nonumber \\
 &&+2\sin 2\theta _{K}\sin \theta _{\ell}\Big[ \cos \phi (
 {Re}{\cal M}_{1}^{0}({\cal M}_{2}^{+*}-{\cal
 M}_{2}^{-*})+{Re}({\cal
 M}_{1}^{+}-{\cal M}_{1}^{-}){\cal M}_{2}^{0*})   \nonumber \\
 &&+\sin \phi ( {Im}{\cal M}_{1}^{0}({\cal M}_{2}^{+*}+ {\cal
 M}_{2}^{-*})-{Im}({\cal M}_{1}^{+}+{\cal M}_{1}^{-}){\cal M}
 _{2}^{0*})\Big]\Big\}  \label{difangle}
 \end{eqnarray}
 with
 \begin{eqnarray}
 |\vec{p}|&=&\sqrt{ E^{\prime 2}- M^{2}_{K^*} }\,,
 \nonumber \\
 E^{\prime }&=&{M^{2}_{B}+M^{2}_{K^*}-q^{2} \over 2M_{B}}\,,
 \nonumber \\
 {\cal M}_{a}^{0} &=&\sqrt{q^{2}}( \frac{E_{K^{*}}}{M_{K^{*}}}
 f_{2}+2t| p_{K^{*}}| \sqrt{q^{2}}\frac{| p_{K^{*}}| }{
 M_{K^{*}}}f_{3})\,,
   \nonumber \\
 {\cal M}_{a}^{\pm } &=&\sqrt{q^{2}}( \pm 2| p_{K^{*}}| \sqrt{
 q^{2}}f_{1}+f_{2}t)\,, \label{M}
 \end{eqnarray}
 where  $a=1(2)$ while $f_{i}=h_{i}$ $(g_{i})$ ($i=1$,
 $2$, $3$). The polarization components ${\cal M}_{a}^{0}$ and
 ${\cal M}_{a}^{\pm }$ in Eq. (\ref{M}) clearly represent the
 longitudinal and transverse polarizations,
 and can be
 easily obtained from Eq. (\ref{ampk1}),
 respectively.
We note that other distributions for
the $K^*$ polarization and CP asymmetries are discussed
in Refs. \cite{Kim} and \cite{Kruger} and
 the photon polarization in $B\to K^*\gamma\to
(K\pi)(l^+l^-)$ is studied in Ref.
\cite{Grossman}.

  From Eqs. (\ref{im1}) and
 (\ref{im2}), we know that ${Im}({\cal M}
 _{i}^{+}-{\cal M}_{i}^{-}){\cal M}_{i}^{0*}$ and ${Im}({\cal M}_{i}^{+}{\cal %
 M}_{i}^{-*})$ are from ${\cal M}_{1\mu }^{(\lambda )}{\cal
 M}_{1\mu ^{\prime
 }}^{(\lambda ^{\prime })\dagger }TrL^{\mu }L^{\mu ^{\prime }}$, while
${Im}%
 {\cal M}_{1}^{0}({\cal M}_{2}^{+*}+{\cal M}_{2}^{-*})-{Im}({\cal M}_{1}^{+}+%
 {\cal M}_{1}^{-}){\cal M}_{2}^{0*}$ is induced by ${\cal M}_{1\mu
 }^{(\lambda )}{\cal M}_{2\mu ^{\prime }}^{(\lambda ^{\prime
 })\dagger }TrL^{\mu }L^{5\mu ^{\prime }}$. Integrating the angular
 dependence in Eq. (\ref{difangle}), we obtain
 \begin{eqnarray*}
 \frac{d\Gamma ( B\rightarrow K^{*} \ell^{+} \ell^{-}\rightarrow
 (K\pi )\ell^{+} \ell^{-}) }{dq^{2}} &=&Br(K^{*}\rightarrow K\pi
 )\frac{\alpha _{em}^{2}G_{F}^{2}| V_{tb}V_{ts}^{*}| ^{2}| \vec{p}|
 }{
 3\times 2^{8}\pi ^{5}M_{B}^{2}} \\
 &&\times [ \sum_{i=1,2}( | {\cal M}_{i}^{0}| ^{2}+| {\cal
 M}_{i}^{+}| ^{2}+| {\cal M}_{i}^{-}|
 ^{2})]
 \\
 &=&\frac{d\Gamma ( B\rightarrow K^{*}\ell^{+} \ell^{-}) }{dq^{2}}
 Br(K^{*}\rightarrow K\pi )\,,
 \end{eqnarray*}
 which conforms the well known equality of
 \[
 Br\left( B\rightarrow K^{*}\ell^{+}\ell^{-}\rightarrow \,K\pi
 \ell^{+} \ell^{-}\right) =Br\left( \,B\rightarrow
 \,K^{*}\,\ell^{+}\,\ell^{-}\right) Br(K^{*}\rightarrow K\pi ).
 \]
 It is interesting to note that
 by integrating out
 $\theta_{\ell}$ and $\phi$ in Eq. (\ref{difangle}),
 we have that
 %the differential decay rates
 %as  a function of $\theta_K$ are
 \begin{eqnarray}
 {d\Gamma \over dq^2 dcos\theta_{K}}&=&\frac{
 G_{F}^{2}\alpha^{2}|\lambda _{t}| ^{2}| \vec{p} | }{2^{10}\pi
 ^{5}m_{B}^{2}}Br(K^{*}\rightarrow K\pi ) \Big\{2\cos ^{2}\theta
 _{K}\sum_{i=1,2}|
 {\cal M}_{i}^{0}| ^{2} \nonumber \\
 && +\sin ^{2}\theta _{K}\sum_{i=1,2}\Big( | {\cal
 M}_{i}^{+}|^{2}+| {\cal M}_{i}^{-}|^{2}\Big)\Big\}\,,
  \label{ltp}
 \end{eqnarray}
which allow us to define
% where the terms associated with $\cos^{2}\theta_{K}$ and
% $\sin^{2}\theta_{K}$ correspond to the longitudinal and transverse
% polarizations of $K^*$, and
% to study the $K^*$ polarization, we may define the
  normalized longitudinal and transverse polarizations of $K^*$ by
 \begin{eqnarray}
 {\cal P}_{L}(q^2)&=& {\sum_{i=1,2}| {\cal M}_{i}^{0}| ^{2} \over
 \sum_{\lambda }\sum_{i}|{\cal
 M}_{i}^{\lambda }|^{2}}, \label{pl}
 \nn\\
 {\cal P}_{T}(q^2)&=& {\sum_{i=1,2}| {\cal M}_{i}^{+}| ^{2}+| {\cal
 M}_{i}^{-}| ^{2} \over \sum_{\lambda }\sum_{i}|{\cal
 M}_{i}^{\lambda }|^{2} }\,,
 \label{PKs}
 \end{eqnarray}
 respectively.

 \subsection{Physical Observables}

  From Eq. (\ref{difangle}), it is clear that there are 9 different
 helicity combinations in the amplitudes.
 As we will show next, among them, 6 are T-even and
 3  T-odd. If each component can be extracted  from
 the angular
 distribution,
 we should have 9 physical observables, which
  can be measured separately
 in $B\rightarrow K^{*}\ell^{+} \ell^{-}$ decays. To archive the
 purpose, we will propose some proper momentum correlation operators, so
 that each component of Eq. (\ref{difangle}) can be singled out and
 measurable experimentally. In the following discussions, we use
  the $K^{*}$ rest frame. The coordinates of relevant momenta are
 choosing as follows:
 \begin{eqnarray}
 p_{B} &=&(\gamma M_{B},0,0,-\gamma \beta M_{B}),\
 p_{\ell^+}=E_{\ell^+}(p_{\ell^+}^{0},\sin \theta _{\ell},0,p_{\ell^+}^{3}),  \nonumber \\
 p_{\ell^+}^{0} &=&\gamma ^{2}( 1+\beta ^{2}-2\beta \cos \theta
 _{\ell}) ,\ p_{\ell^+}^{3}=\gamma ^{2}( ( 1+\beta ^{2}) \cos
 \theta
 _{\ell}-2\beta ) ,  \nonumber \\
 \beta &=&\frac{| \vec{p}| }{E^{\prime }},\ \gamma =\frac{1
 }{\sqrt{1-\beta ^{2}}},  \nonumber \\
 E_{K} &=&\omega _{K}={\frac{M_{K^{*}}}{2}},\ E_{\ell^+}=\omega
 _{\ell^{+}}={\frac{ \sqrt{q^{2}}}{2}}\,,
   \label{frame}
 \end{eqnarray}
 where $\beta $ and $\gamma $ are the usual Lorentz transformation
 factors. From
 Eq. (\ref{frame}), we have
 \begin{eqnarray}
 \sin \theta _{\ell} &=&\frac{\vec{p}_{\ell^{+}}\times
 \vec{p}_{B}}{|\vec{p} _{B}|\omega _{\ell^{+}}}\,,
   \nonumber \\
 \sin \theta _{K} &=&\frac{\vec{p}_{B}\times \vec{p}_{K}}{|\vec{p}_{B}|\omega
 _{K}} \,,
  \nonumber \\
 \cos \phi &=&\frac{( \vec{p}_{B}\times \vec{p}_{K}) \cdot (
 \vec{p}_{\ell^{+}}\times \vec{p}_{B} ) }{| \vec{p}_{B}\times
 \vec{p} _{K} | | \vec{p}_{\ell^{+}}\times \vec{p}_{B}| }\,,
   \nonumber \\
 \sin \phi &=&| \vec{p}_{B}| \frac{\vec{p}_{K}\cdot ( \vec{p}
 _{B}\times \vec{p}_{\ell^{+}} ) }{| \vec{p}_{B}\times \vec{p}
 _{K}| | \vec{p}_{B}\times \vec{p}_{\ell^{+}}| }\,.
   \label{angles}
 \end{eqnarray}
 To relate the above angles to those in  Eq.
(\ref{angles}),
 we use  momentum correlations denoted by ${\cal O}_i$ and
 define the physical observables by \be \langle {\cal O}_{i}\rangle
 &\equiv & \int {\cal O}_{i}\omega _{i}(u_{\theta _{K}},v_{\theta
 _{\ell}}){\frac{d\Gamma }{dq^{2}}} \ee where $\omega
 _{i}(u_{\theta _{K}},v_{\theta _{\ell}})$ are sign functions,
 $i.e.$, \be \omega _{i}(u_{\theta _{K}},v_{\theta
 _{\ell}})&=&{\frac{u_{\theta _{K}}v_{\theta_{\ell}}}{|u_{\theta
 _{K}}v_{\theta _{\ell}}|}} \ee with $u_{\theta _{i}}$ being $\sin
 \theta _{i}$ or $\cos \theta _{i}$. The asymmetries ${\cal A}_{i}$
 and statistical significances $\varepsilon _{i}$ of ${\cal O}_{i}$
 are given by
 \begin{eqnarray}
 {\cal A}_{i}(q^{2}) &=&{\frac{<{\cal O}_{i}>}
 {\int {\frac{d\Gamma }{dq^{2}}}}}\,,
   \nonumber \\
 \varepsilon _{i}(q^{2}) &=&{\frac{<{\cal O}_{i}>}{\sqrt{\int {\frac{d\Gamma
 }{dq^{2}}}\cdot \int {\cal O}_{i}^{2}{\frac{d\Gamma }{dq^{2}}}}}}\,.
 \label{Aid}
 \end{eqnarray}
 We can also define the integrated asymmetries and statistical
 significances by
 \begin{eqnarray}
 \bar{{\cal A}}_{i} &=&{\frac{\int {\cal O}_{i}\omega
 _{i}(u_{\theta _{K}},v_{\theta _{\ell}})d\Gamma }{\int d\Gamma
 }}\,,
   \nonumber \\
 \bar{\varepsilon}_{i} &=&{\frac{\int {\cal O}_{i}\omega
 _{i}(u_{\theta
 _{K}},v_{\theta _{\ell}})d\Gamma }{\sqrt{\int d\Gamma \cdot \int {\cal O}%
 _{i}^{2}d\Gamma }}}\,.
 \end{eqnarray}
 The numbers of $B$ mesons required to observe the effects at the
 $n\,\sigma $ level are given by
 \be
 N_{i}&=&{\frac{n^2}{BR\cdot A_{i}^{2}}}\,,
 \ee
  where $A_i$ represent the asymmetries or the statistical significances
 %${\cal A}_i$ or $\varepsilon_i$
 and $BR$ the branching ratios of
 $B\to K\pi \ell^{+} \ell^{-}$.

 To study the various parts of the angular distributions in Eq.
 (\ref{difangle}), we use  nine operators ${\cal O}_i\
 (i=1,2,\cdots, 9)$ and sign functions $\omega_i$ as follows: \be
 {\cal O}_{1} &=&4\frac{| \vec{p}_{\ell^{+}}\times \vec{p}_{B}|
 ^{2}} {| \vec{p}_{B}| ^{2}\omega _{\ell^{+}}^{2}}-3\frac{|
 \vec{p}_{B}\times \vec{p}_{K}| ^{2}}{| \vec{p}_{B}| ^{2}\omega
 _{K}^{2}}\,,
 \nn\\
 \omega_1&=&\omega_1(\sin\theta_K,\sin\theta_{\ell})\,,
 \label{opr1}
 \\
 \nn\\
 {\cal O}_{2} &=&2\frac{| \vec{p}_{B}\times \vec{p}_{K}| ^{2}}{ |
 \vec{p}_{B}| ^{2}\omega _{K}^{2}}-\frac{| \vec{p}
 _{\ell^{+}}\times \vec{p}_{B}| ^{2}}{| \vec{p}_{B}| ^{2}\omega
 _{\ell^+}^{2}} \,,
 \nn\\
  \omega_2&=&\omega_1\,,
 \label{opr2}
 \\
 \nn\\
 {\cal O}_{3}&=&\frac{( \vec{p}_{B}\times \vec{p}_{K}) \cdot (
 \vec{p}_{\ell^{+}}\times \vec{p}_{B}) }{| \vec{p}_{B}\times
 \vec{p}_{K}| | \vec{p}_{\ell^{+}}\times \vec{p}_{B}| }\,,
 \nn\\
  \omega_3&=&\omega_3(\cos\theta_K,\cos\theta_{\ell})\,,
 \label{opr3}
 \\
 \nn \\
 {\cal O}_{4}&=&\frac{[ ( \vec{p}_{B}\times \vec{p}_{K}) \cdot (
 \vec{p}_{\ell^{+}}\times \vec{p}_{B}) ] ^{2}}{| \vec{p}_{B}\times
 \vec{p}_{K}| ^{2}| \vec{p}_{\ell^{+}}\times \vec{p} _{B}| ^{2}}-|
 \vec{p}_{B}| ^{2}\frac{( \vec{p} _{B}\cdot
 \vec{p}_{\ell^{+}}\times \vec{p}_{K}) ^{2}}{| \vec{p} _{B}\times
 \vec{p}_{K}| ^{2}| \vec{p}_{\ell^{+}}\times \vec{p} _{B}| ^{2}}\,,
 \nn\\
  \omega_4&=&\omega_1\,,
 \label{opr4}
  \\
 \nn\\
  {\cal O}_{5}&=& 1\,,
 \nn\\
  \omega_5&=&\omega_5(\sin\theta_K,\cos\theta_{\ell})\,,
 \label{opr5}
 \\
 \nn\\
  {\cal O}_6&=& {\cal O}_{3}\,,
 \nn\\
  \omega_6&=&\omega_6(\cos\theta_K,\sin\theta_{\ell})\,,
 \label{opr6}
  \\
 \nn\\
 {\cal O}_{7}&=&| \vec{p}_{B}| \frac{\vec{p}_{K}\cdot (
 \vec{p}_{B}\times \vec{p}_{\ell^{+}}) }{| \vec{p}_{B}\times
 \vec{p} _{K}| | \vec{p}_{B}\times \vec{p}_{\ell^{+}}| } \,,
 \nn\\
  \omega_7&=&\omega_7(\cos\theta_K,\cos\theta_{\ell})\,,
 \label{opr7}
 \\
 \nn\\
 {\cal O}_{8}& =&| \vec{p}_{B}| \frac{( \vec{p}_{B}\cdot
 \vec{p}_{\ell^{+}}\times \vec{p}_{K}) ( \vec{p}_{B}\times \vec{p}
 _{K}) \cdot ( \vec{p}_{\ell^{+}}\times \vec{p}_{B}) }{|
 \vec{p}_{B}\times \vec{p}_{K}| ^{2}| \vec{p}_{\ell^{+}}\times
 \vec{p} _{B}| ^{2}} \,,
 \nn\\
  \omega_8&=&\omega_1\,,
 \label{opr8}
 \\
 \nn\\
 {\cal O}_{9}&=&{\cal O}_{7}\,,
 \nn\\
  \omega_9&=&\omega_6\,.
 \label{opr9} \ee It is clear that the first six operators ${\cal
 O}_{i}\ (i=1-6)$ in Eqs. (\ref {opr1})-(\ref{opr6}) are T-even
 observables, whereas the last three ${\cal O} _{j}\ (i=7-9)$ T-odd
 ones. We remark that the operators and sign functions
in Eqs. (\ref {opr1})-(\ref{opr9}) are the simplest ones to discuss the
momentum correlations.

  From Eqs. (\ref{difangle}),
 (\ref{Aid}), and
 (\ref{opr1})-(\ref{opr9}), we find that

 \begin{eqnarray}
 {\cal A}_1(q^2)
  &=&4\cdot {\frac{32\pi }{9\Gamma _{0}}}\sum_{i=1,2}|{\cal M}
 _{i}^{0}|^{2} \,,
 \nn\\
 {\cal A}_2(q^2)
  &=&{\frac{64\pi }{9\Gamma _{0}}}\sum_{i=1,2}(|{\cal M}_{i}^{+}|^{2}
 +|{\cal M}_{i}^{-}|^{2})\,,
 \nn \\
 {\cal A}_3(q^2) &=&-{\frac{16\pi }{9\Gamma
 _{0}}}\sum_{i=1,2}Re({\cal M}_{i}^{+}+{\cal M}_{i}^{-}){\cal
 M}_{i}^{0*} \,,
 \nn\\
 {\cal A}_4(q^2) &=&-2\cdot {\frac{16\pi }{9\Gamma
 _{0}}}\sum_{i=1,2}Re({\cal M}_{i}^{+} {\cal M}_{i}^{-*})\,,
 \nn \\
 {\cal A}_5(q^2)
  &=&-2\cdot {\frac{8\pi }{9\Gamma _{0}}}( 2Re{\cal M}_{1}^{+}{\cal
 M}_{2}^{+*}-2Re{\cal M}_{1}^{-}{\cal M}_{2}^{-*}) \,,
 \nn \\
 {\cal A}_6(q^2)
  &=&2\cdot {\frac{2\pi ^{2}}{3\Gamma _{0}}}( Re{\cal M}_{1}^{0}(
 {\cal M}_{2}^{+*}-{\cal M}_{2}^{-*})+Re({\cal M}_{1}^{+}-{\cal
 M}_{1}^{-}) {\cal M}_{2}^{0*})\,,
 \nn\\
 {\cal A}_7(q^2) &=&-{\frac{16\pi }{9\Gamma
 _{0}}}\sum_{i=1,2}Im({\cal M}_{i}^{+}-{\cal M}_{i}^{-}){\cal
 M}_{i}^{0*}\,,
 \nn \\
 {\cal A}_8(q^2) &=&-2\cdot {\frac{8\pi }{9\Gamma
 _{0}}}\sum_{i=1,2}Im({\cal M}_{i}^{+} {\cal M}_{i}^{-*})\,,
 \nn\\
 {\cal A}_9(q^2)
  &=&2\cdot {\frac{2\pi ^{2}}{3\Gamma _{0}}}( Im{\cal M}_{1}^{0}(
 {\cal M}_{2}^{+*}+{\cal M}_{2}^{-*})-Im({\cal M}_{1}^{+}+{\cal
 M}_{1}^{-}) {\cal M}_{2}^{0*})\,, \label{Ai}
 \end{eqnarray}
 where $\Gamma _{0}=64\pi /9\sum_{\lambda }\sum_{i}|{\cal M}_{i}^{\lambda
 }|^{2}$.
 We note that the asymmetry ${\cal A}_{1 (2)}$
 in Eq. (\ref{Ai}) is related to
  the longitudinal (transverse) polarization of $K^*$ in Eq. (\ref{PKs}).
 We can also evaluate $\varepsilon_i(q^2)$ similar to those
 in Eq. (\ref{Ai}) except the denominators due to
 $\int {\cal O}_i^2d\Gamma/dq^2$.

 \section{Numerical Analysis}
 In our numerical analysis,
 we use $f_B=0.19$, $f_{K^*}=0.21$, $f^{T}_{K^*}=0.17$, $M_{B}=5.28$,
 $M_{K^*}=0.892$, $m_{b}=4.8\ GeV$, $\lambda_t=0.04$,
 $\alpha_{em}=1/129$,
 and $c=0.4$, and
 we take the $B$ meson wave function as
 \begin{equation}
 \phi _{B}(x,b)=N_{B}x^{2}(1-x)^{2}\exp \Big[ -\frac{1}{2}(
 \frac{xM_{B} }{\omega _{B}}) ^{2}-\frac{\omega
 _{B}^{2}b^{2}}{2}\Big] , \label{bwf}
 \end{equation}
 where $\omega _{B}$ is the shape parameter \cite{BW} and $N_{B}$
 is determined by the normalization of the $B$ meson wave function,
 given by
 \begin{eqnarray}
 \int_{0}^{1}dx\phi_{B}(x,0) &=& {f_B \over 2\sqrt{2N_{c}}}.
 \end{eqnarray}
 Since the PQCD  can be only applied to the outgoing particle of carrying
 a large energy, where a small coupling constant $\alpha_{s}$ expansion
 is reliable, we only perform our numerical analysis
  in $q^{2}\leq 10\ GeV^2$.

 \subsection{form factors}
 In Table \ref{vff},
 \begin{table}
 \caption{Form factors for $B\to K^*$ at $q^2=0$ in various QCD models.}
 \label{vff}
 \begin{center}
 \begin{tabular}{lllllll}
 \hline Model & \multicolumn{1}{c}{$V( 0)$} & $A_{0}( 0) $ &
 \multicolumn{1}{c}{$A_{1}(
 0) $} & $A_{2}( 0) $ & \multicolumn{1}{c}{$T_{1}( 0) $} & $%
 T_{3}( 0) $
 \\ \hline
 LEET \cite{Burdman} &
 \multicolumn{1}{c}{$0.36\pm 0.04$} & \multicolumn{1}{c}{} &
 \multicolumn{1}{c}{$0.27\pm 0.03$} & \multicolumn{1}{c}{} &
 \multicolumn{1}{c}{$0.31\pm 0.02$} & \multicolumn{1}{c}{} \\
 \hline
 QM \cite{MS} & \multicolumn{1}{c}{$0.44$} &
 \multicolumn{1}{c}{$0.45$} &
 \multicolumn{1}{c}{$0.36$} & \multicolumn{1}{c}{$0.32$} & \multicolumn{1}{c}{%
 $0.39$} & \multicolumn{1}{c}{$0.27$} \\ \hline
  LCSR \cite{Ali} & \multicolumn{1}{c}{$0.399$} &
 \multicolumn{1}{c}{$0.412$} & \multicolumn{1}{c}{$0.294$} &
 \multicolumn{1}{c}{$0.246$} & \multicolumn{1}{c}{$ 0.334$} &
 \multicolumn{1}{c}{$0.234$}
 \\ \hline LFQM \cite{LF} & \multicolumn{1}{c}{$0.35$}
 &\multicolumn{1}{c}{$0.32$} & \multicolumn{1}{c}{$0.26$}
 &\multicolumn{1}{c}{$0.23$}
 &\multicolumn{1}{c}{$0.32$}&\multicolumn{1}{c}{$0.21$} \\ \hline
 PQCD (I) & \multicolumn{1}{c}{$0.355$} &
 \multicolumn{1}{c}{$0.407$} & \multicolumn{1}{c}{$0.266$} &
 \multicolumn{1}{c}{$0.202$} & \multicolumn{1}{c}{$0.315$} &
 \multicolumn{1}{c}{$0.207$} \\
 \multicolumn{1}{r}{(II)} & \multicolumn{1}{c}{$0.332$} & \multicolumn{1}{c}{$%
 0.381$} & \multicolumn{1}{c}{$0.248$} &
 \multicolumn{1}{c}{$0.189$}
 & \multicolumn{1}{c}{$0.294$} & \multicolumn{1}{c}{$0.193$} \\
 \hline
 \end{tabular}
 \end{center}
 \end{table}
 we show
 the form factors
 parametrized in Eq. (\ref{ff}) with (I) $\omega_{B}=0.40\ GeV$ and
 (II) $\omega_{B}=0.42\ GeV$ at $q^{2}=0$.
  As comparisons,
 in the table we also give results from
 %other QCD approaches, such as
 the light cone sum rule (LCSR) \cite{Ali},
 quark model (QM) \cite{MS}, and light front quark model (LFQM)
 \cite{LF}.
 Since in  the
 large energy effective theory (LEET) seven independent form
 factors in Eq. (\ref{ff}) can be reduced to two in the small
 $q^{2}$ region \cite{Charles}, in Table \ref{vff}, we only show
 $T_{1}(0)$, $V(0)$, and $A_{1}(0)$ \cite{Burdman} extracted by
 combining the LEET and the experimental data on $B\rightarrow K^*
 \gamma$.

In our
 following numerical analysis, we only take the minimal results in the
LCSR, which are consistent with those from the extraction of the LEET,
 as the representation of
 the LCSR. From Table \ref{vff}, we find  that our results from the
 PQCD agree with those from all other models
 except the QM.

 It is interesting to point out that the decay branching ratio (BR)
 of  $B\rightarrow \phi K^{0*}$ is found to be $1.7\ (1.5)\times
 10^{-5}$ for $\omega_{B}=0.40\ (0.42)$ \cite{CKLVV}, comparing
 with the recent BELLE's result of
 $1.3^{+0.64}_{-0.52}\pm 0.21 \times 10^{-5}$ \cite{BELLEVV}.
 Here,
 the overwhelming contributions to BRs of
 $B\rightarrow \phi K^*$  are from the longitudinal parts, where
  the form factor $A_{0}$  plays an essential role.
 We remark that $A_{0}$  does not appear in our present analysis
 due to the light lepton mass neglected. However, one can obtain the
 value of
 $A_{0}$  if more accurate measurements on the modes of
 $B\rightarrow \phi K^*$
  are available in the near future. After getting $A_{0}(0)$
 and $A_{1}(0)$ ,
we can find $A_{2}(0)$ from the identity in Eq. (\ref{idff}).
 Furthermore, by using the relations
 among the form factors in the HQET \cite{Burdman}, one can
 easily get $T_{3}(0)$ as well. In sum, in terms of the
 measurements of $B\rightarrow \phi K^*$ and $B\rightarrow K^*
 \gamma$ together with the HQET and LEET, all form factors at
 $q^2=0$ for $B\rightarrow K^*$  can be extracted
 model-independently.

\begin{figure}[tbp]
 \vspace{1cm} \centerline{ \psfig{figure=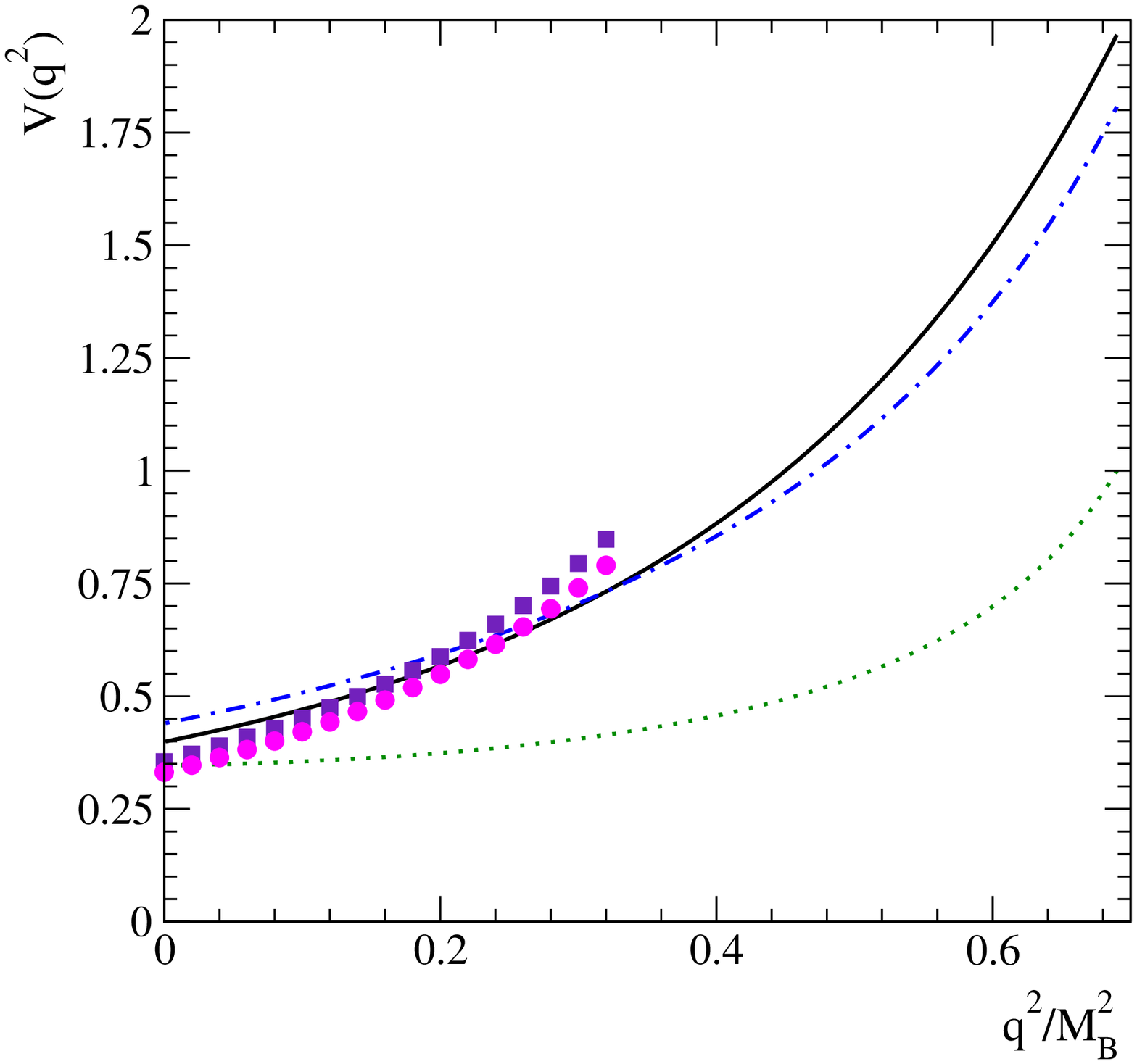,height=3.in }  }
 \caption{Form factor for $V(q^2)$ as a fuction of $q^2/M_B^2$. The
 curve with squares (circles) stands for the PQCD calculation
  with $\omega_{B}=0.40\ (0.42)$ and the solid curve denotes
 the minimal values in the LCSR \cite{Ali}, while the
 dash-dotted and dotted ones represent the
 results of the QM and LFQM, respectively.  }
 \label{figv}
 \end{figure}

\begin{figure}[tbp]
 \vspace{1cm} \centerline{ \psfig{figure=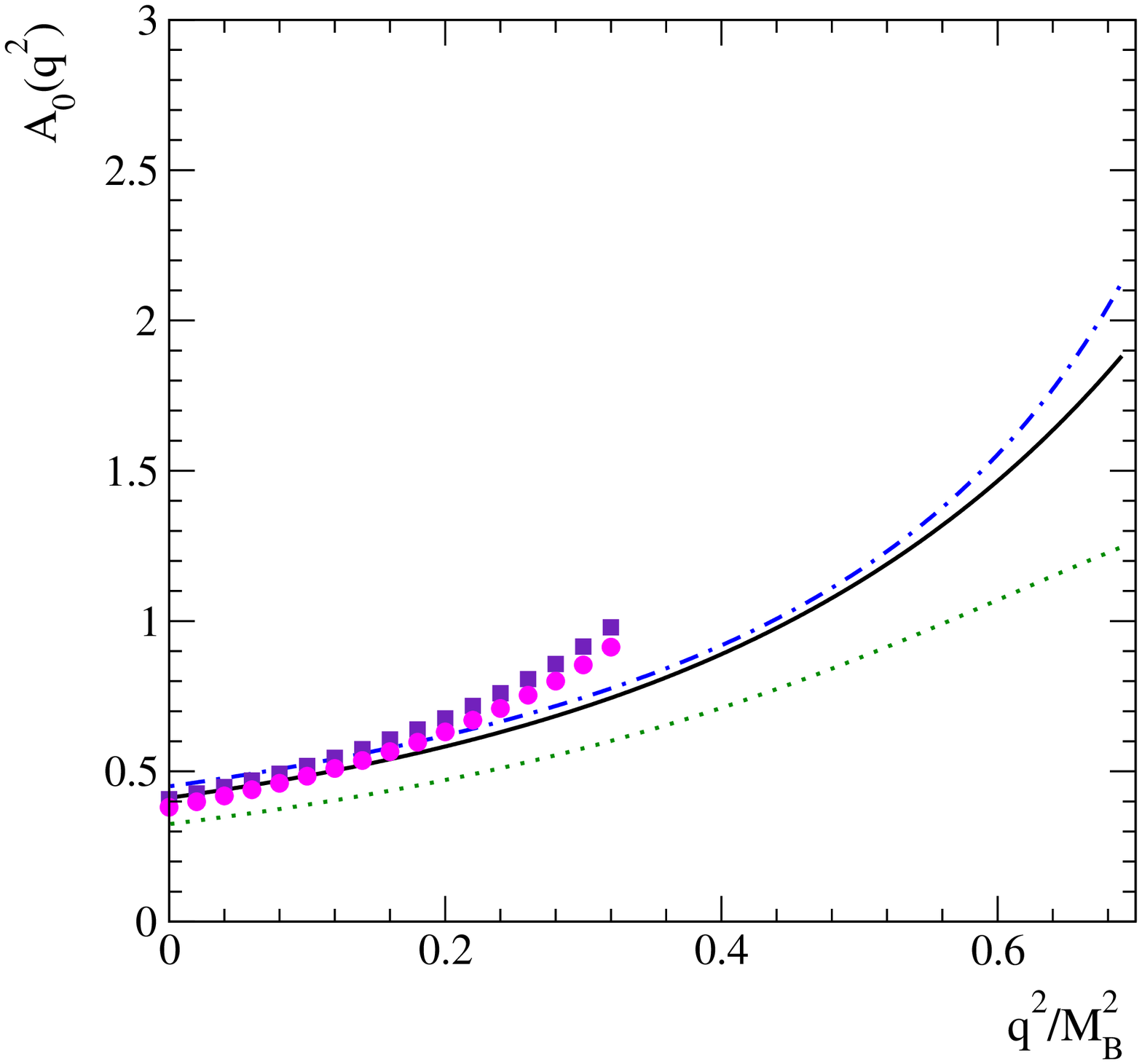,height=3.in }  }
 \caption{Same as Figure \ref{figv} but for $A_{0}(q^2)$.}
 %$A_{0}(q^2)$ Legend is the same as figure \ref{figv} }
 \label{figa0}
 \end{figure}

\begin{figure}[tbp]
 \vspace{1cm} \centerline{ \psfig{figure=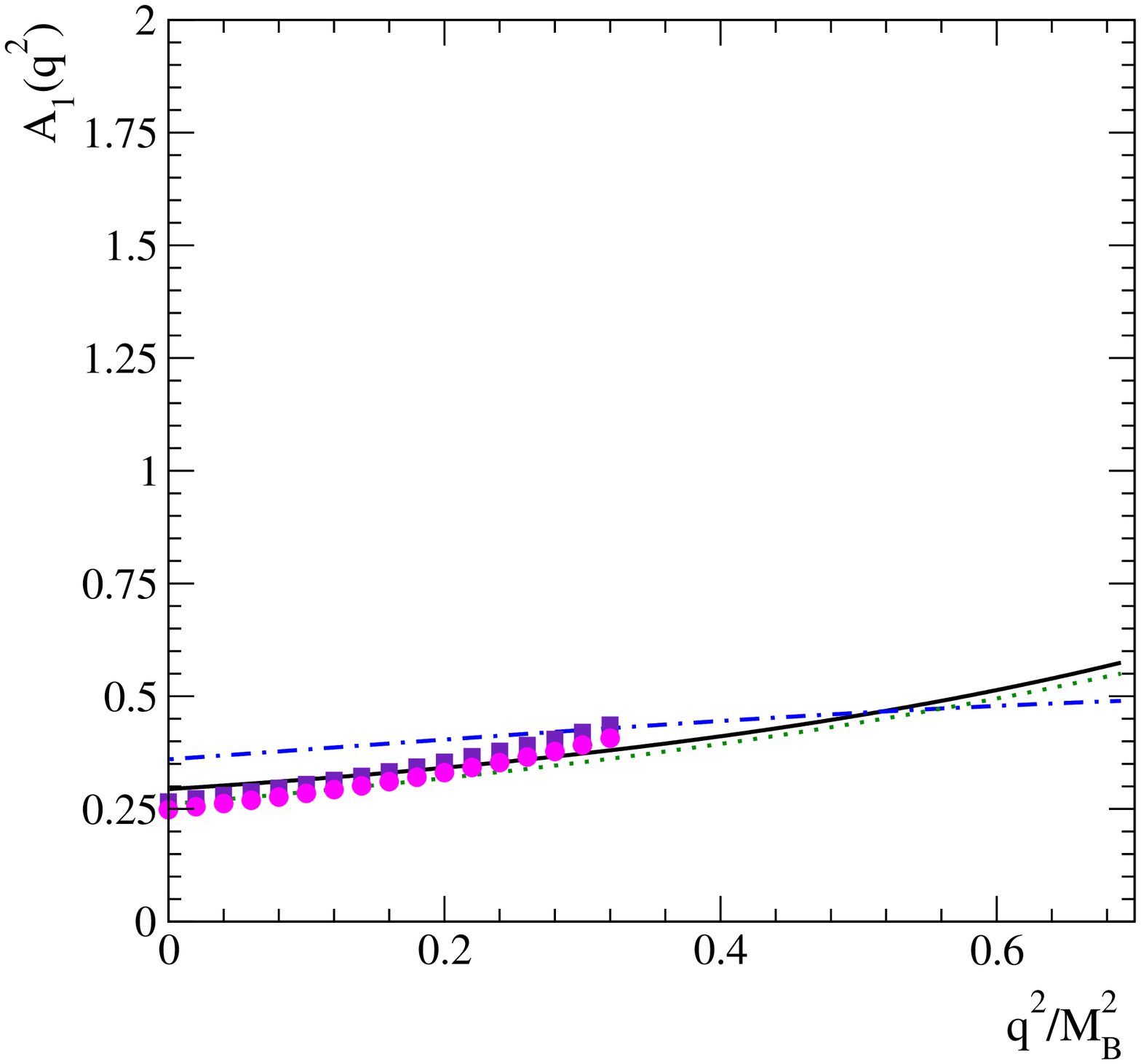,height=3.in }  }
 \caption{Same as Figure \ref{figv} but for $A_{1}(q^2)$.}
 %$A_{1}(q^2)$ form factor. Legend is the same as figure \ref{figv}}
 \label{figa1}
 \end{figure}

\begin{figure}[tbp]
 \vspace{1cm} \centerline{ \psfig{figure=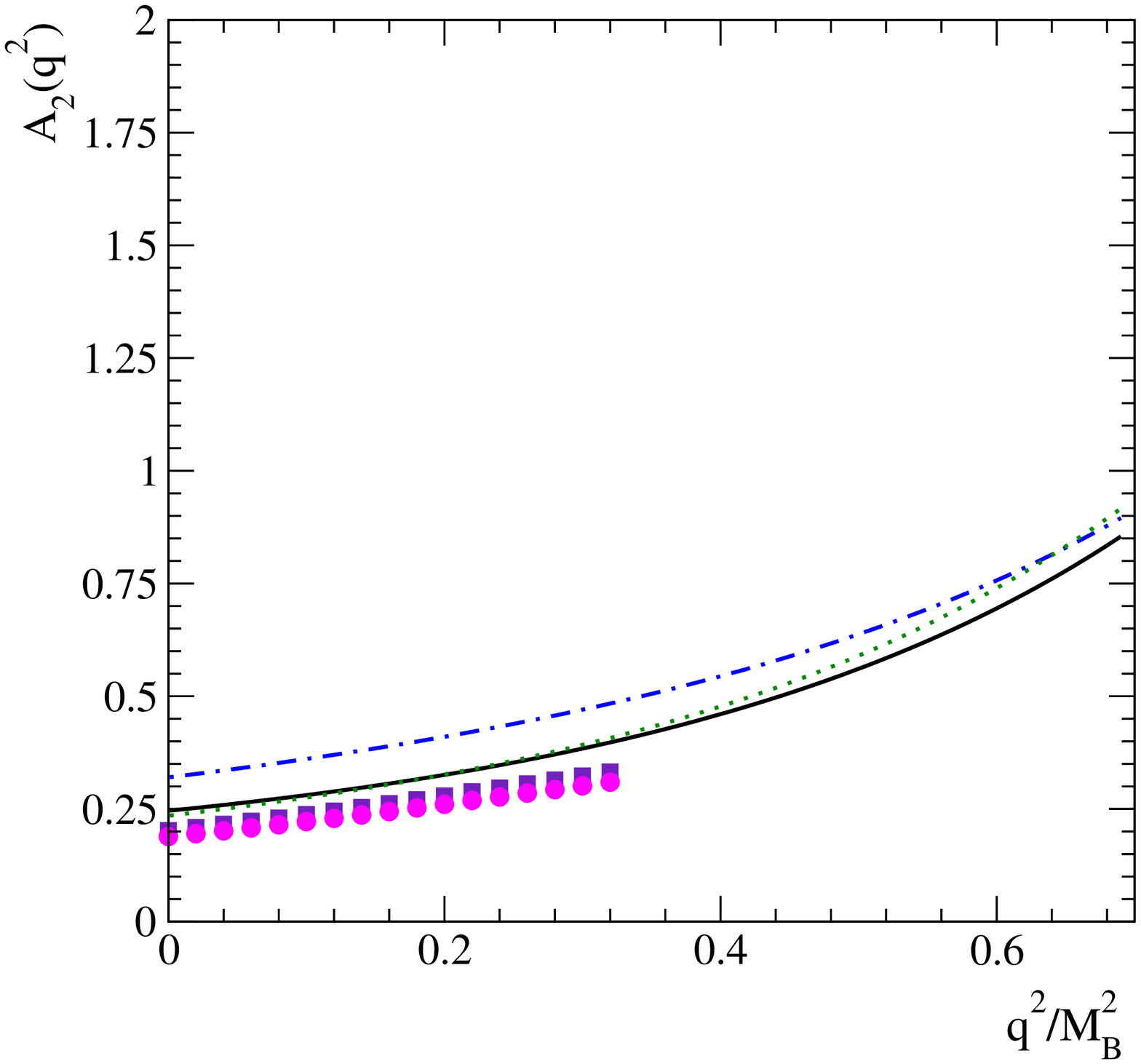,height=3.in }  }
 \caption{Same as Figure \ref{figv} but for $A_{2}(q^2)$.}
 % form factor. Legend is the same as figure \ref{figv} }
 \label{figa2}
 \end{figure}

\begin{figure}[tbp]
 \vspace{1cm} \centerline{ \psfig{figure=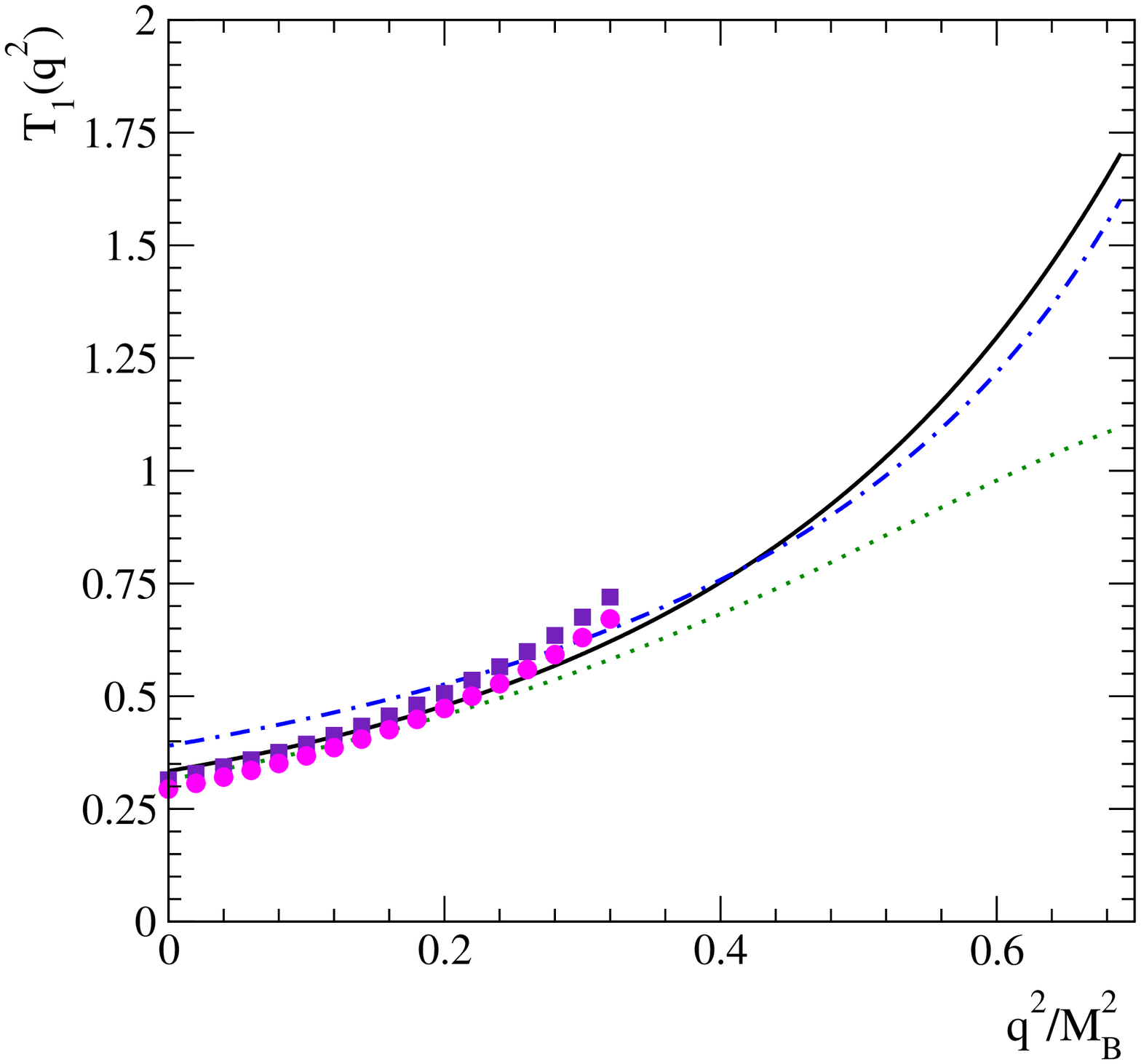,height=3.in }  }
 \caption{Same as Figure \ref{figv} but for $T_{1}(q^2)$.}
 % form factor. Legend is the same as figure \ref{figv} }
 \label{figt1}
 \end{figure}

\begin{figure}[tbp]
 \vspace{1cm} \centerline{ \psfig{figure=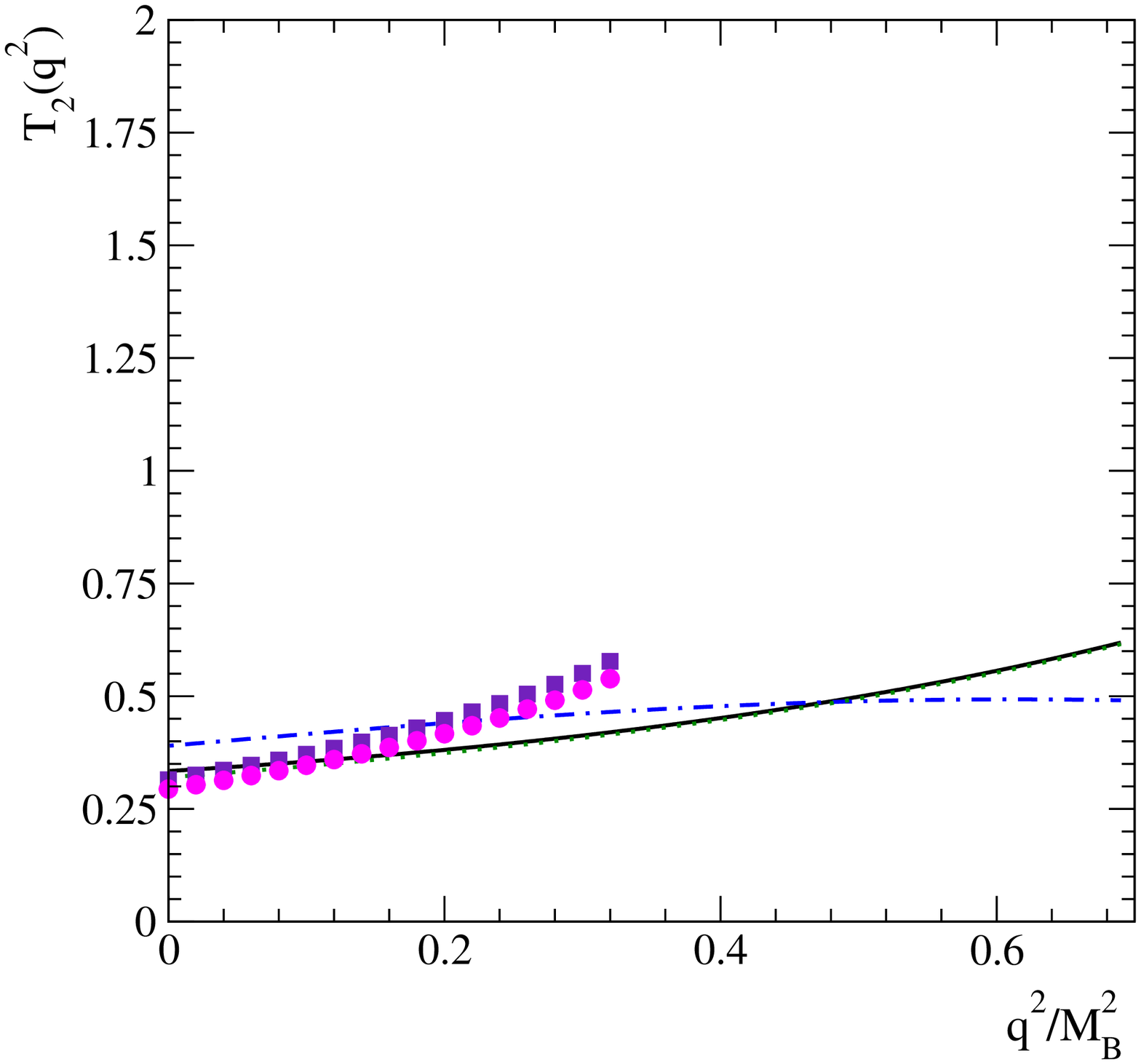,height=3.in }  }
 \caption{Same as Figure \ref{figv} but for $T_{2}(q^2)$.}
 % form factor. Legend is the same as figure \ref{figv} }
 \label{figt2}
 \end{figure}

\begin{figure}[tbp]
 \vspace{1cm} \centerline{ \psfig{figure=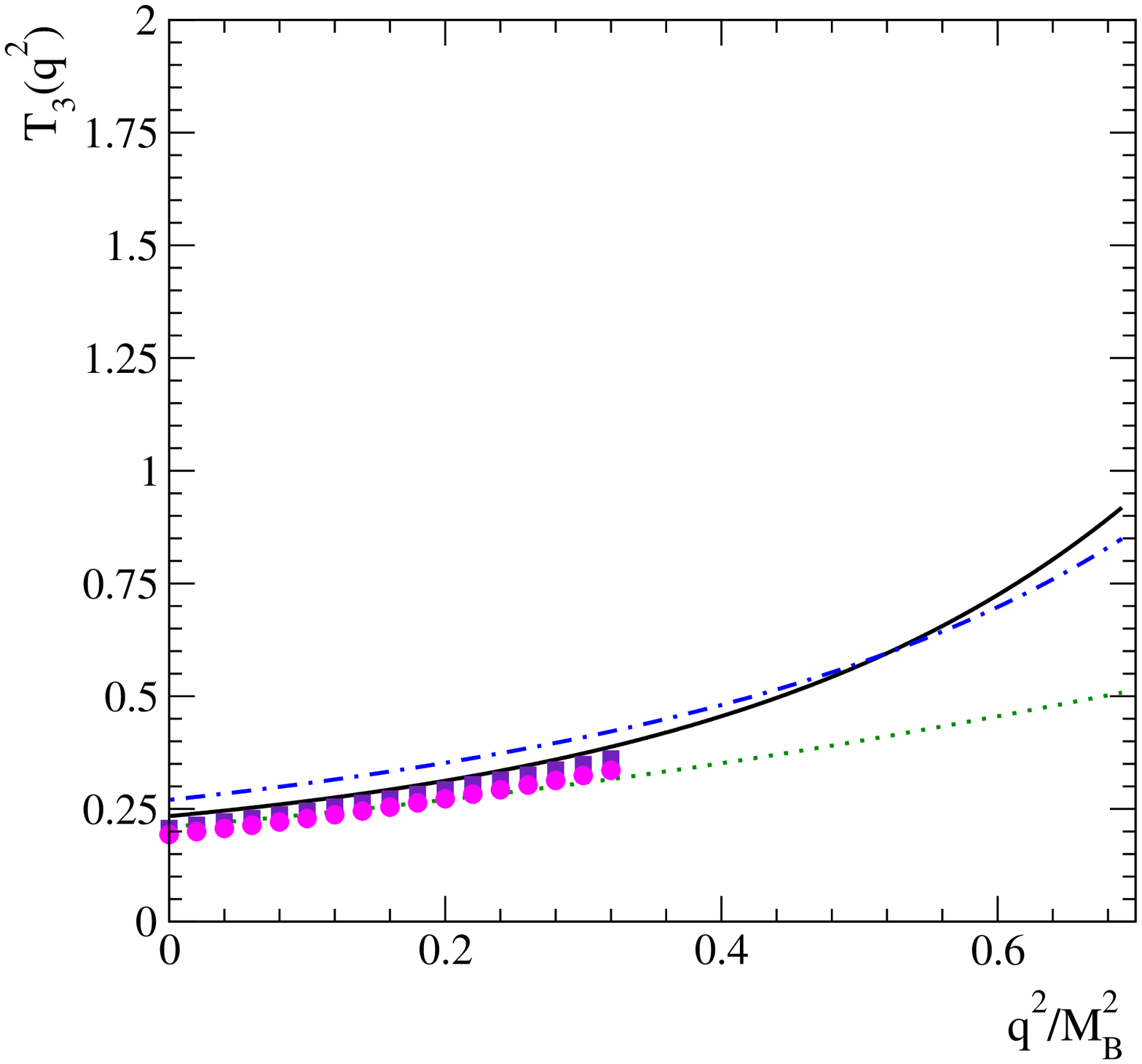,height=3.in }  }
 \caption{Same as Figure \ref{figv} but for $T_{3}(q^2)$.}
 % form factor. Legend is the same as figure \ref{figv} }
 \label{figt3}
 \end{figure}

 In Figures \ref{figv}-\ref{figt3}, we display the form factors
 $V(q^2)$, $A_{0,1,2}(q^2)$, and $T_{1,2,3}(q^2)$ as functions of
 $s=q^2/M_B^2$, in the LCSR, QM, LFQM, PQCD (I), and PQCD (II),
 representing by
 the solid, dash-dotted, dotted, square and circle curves,
 respectively.
 %The curves with squares  (circles) denote the PQCD
 %calculations with $\omega=0.40$ ($0.42$).

 %
 \subsection{Differential decay rates}
\begin{figure}[tbp]
 \vspace{1cm} \centerline{ \psfig{figure=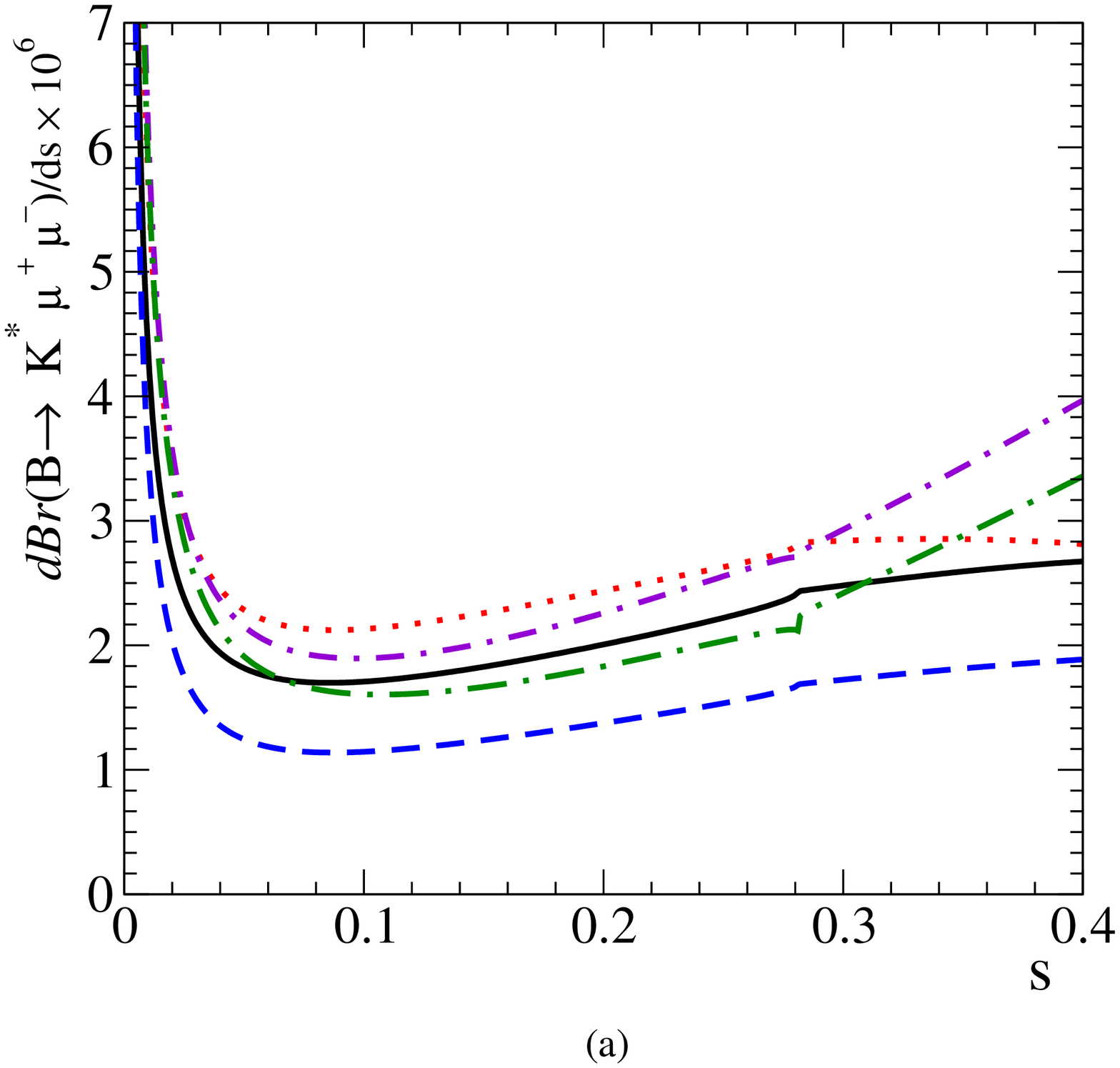,height=3.in }
 \psfig{figure=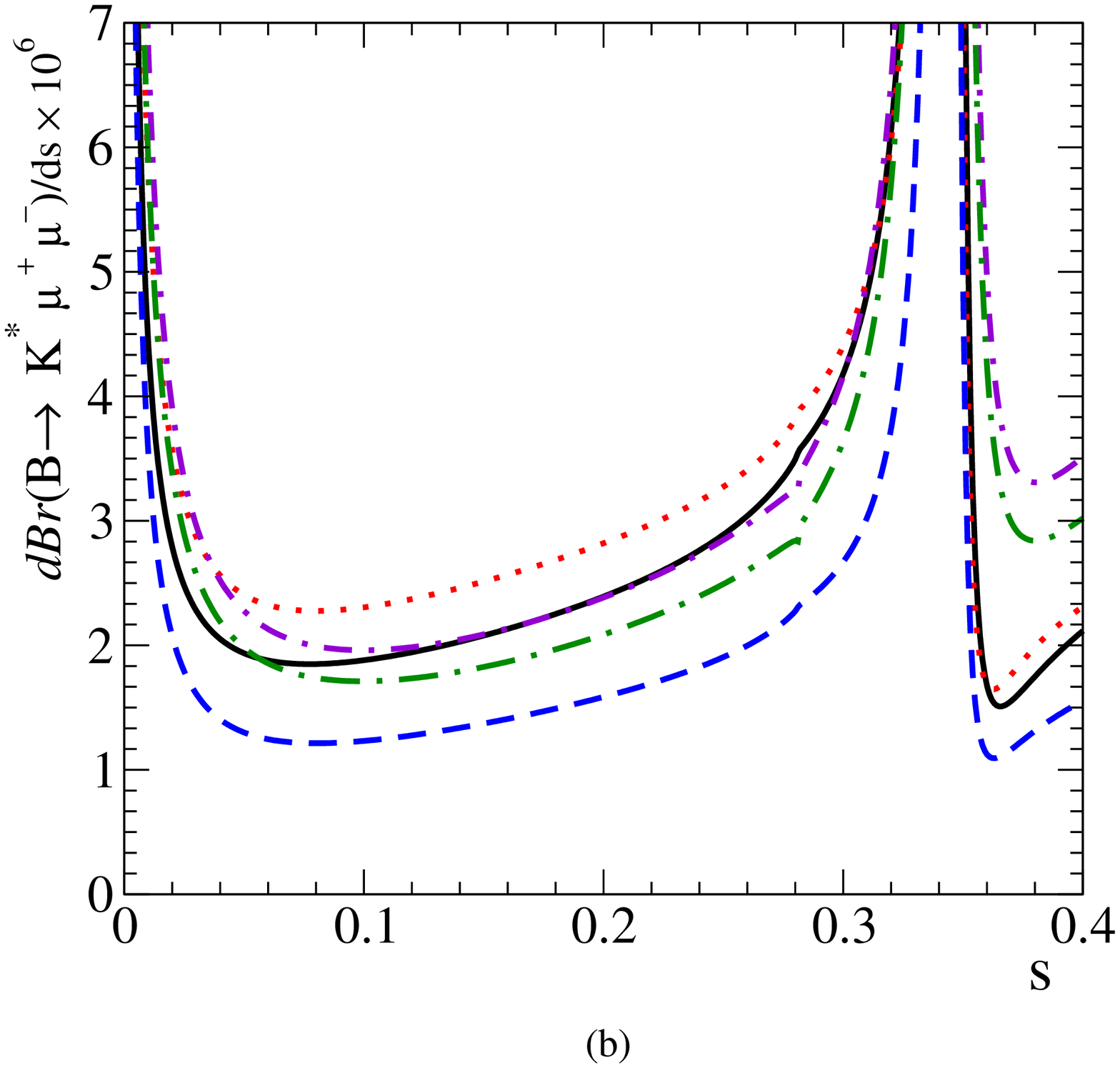,height=3.in }}
 \caption{Differential decay rate of $B\to K^*\mu^+\mu^-$
 as a function of $s=q^2/M^2_{B}$. The solid, dotted, and
 dashed curves stand for the results of the LCSR, QM, and LFQM;
  the upper and
 lower dash-dotted ones are those from the PQCD (I) and (II),
and (a) and (b) represent the results
 with and without resonant effects, respectively.
}
 \label{figdifbr}
 \end{figure}

\begin{figure}[tbp]
 \vspace{0cm} \centerline{ \psfig{figure=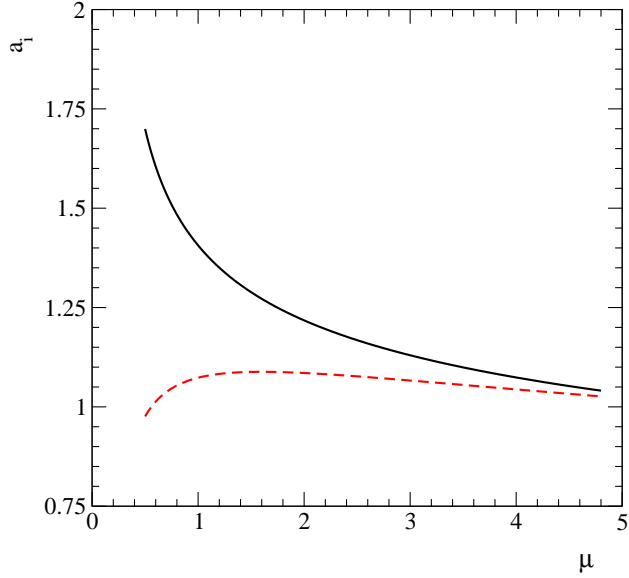,height=3.in } }
 \caption{Wilson coefficients as a function of $\mu$ normalized by
 themselves at $\mu=m_b$. The solid and dashed curves are for
 $a_7=C_7(\mu)/C_7(m_b)$  and $a_9=C_9(\mu)/C_9(m_b)$,
 respectively.} \label{figc7c9}
 \end{figure}

\begin{figure}[tbp]
 \vspace{1.cm} \centerline{ \psfig{figure=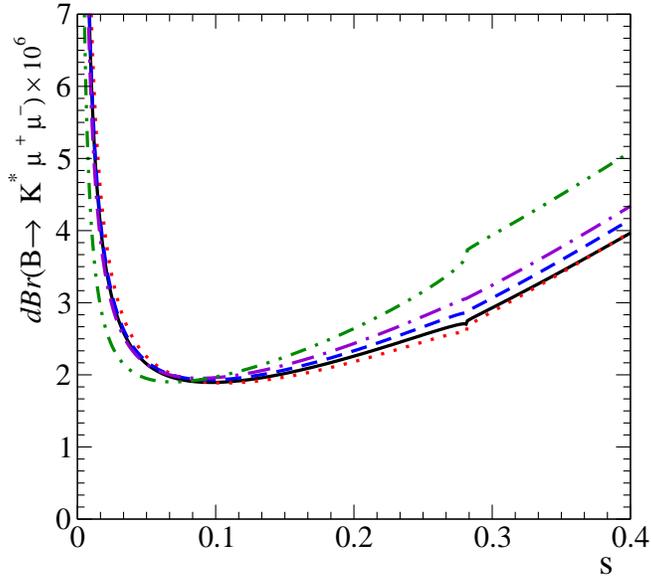,height=3.in } }
 \caption{ Differential decay rate of $B\to K^*\mu^+\mu^-$ for
 different $\mu$ scales. The dotted, dashed, dash-dotted,
 double-dot-dashed curves stand for $\mu=1.3$, $1.5$, $1.7$, and
 $5.0$ GeV, respectively, while the solid one expresses the result
 of the full PQCD formalism.} \label{figmu}
 \end{figure}

 We now present the  dilepton invariant mass distributions for $B\to
 K^* \ell^+ \ell^-$
  by integrating all angular dependence in the phase
 space. The distribution for the decay mode with a muon pair in
 various QCD approaches is shown in Figure \ref{figdifbr},
  where  (a) and (b) represent the results
 with and without resonant effects, respectively.
 %In the figures,
 %we have only shown the minimal values (solid curves) in the LCSR.
  From the figures, we find that the PQCD results are consistent
 with the minimal ones in the LCSR approach due to similar form
 factors in the lower $q^2$ region in both models. Here,  we have
 set that all WCs are involved at the $m_{b}$ scale
 for all QCD approaches except the PQCD one. As emphasized in Sec.
 III, in the PQCD formalism WCs should be
 convoluted with hard parts and meson wave functions of $B$ and $K^*$.
 Due to the hard gluon exchange, with the momentum squared,
 $|k_{\bot}|^2$, being off-shellness in magnitude of
 $O(\bar{\Lambda}M_{B})$, dominated in the fast recoil region, the
 PQCD approach involves a lower scale \cite{Chen1,KLS1}. As a
 consequence, even using the concept of the naive factorization,
 where the decay amplitude is expressed by the product of the
 WC and the corresponding form factor, the typical
 scale $t_{0}$ in the PQCD should be much less than $M_{B}$ or
 $M_{B}/2$. To illustrate the scale dependent on the WCs,
 we display the $C_{7}(\mu)$ and $C_{9}(\mu)$, renormalized
 by themselves at $m_{b}$, as functions of $\mu$-scale
 in Figure \ref{figc7c9}.
  In Figure \ref{figmu}, we show the decay rate of $B\to K^*\mu^+\mu^-$
 with the
 relevant WCs fixed at $t_{0}=1.3$, $1.5$, $1.7$ and
 $5.0$ GeV, respectively.
  From the figure, we find that
 the result with
  $t_{0}=1.3$ GeV is compatible with that from the formal PQCD
 approach. Similar conclusion is also expected for the electron mode.

 \subsection{Physical observables}

 Because the numerical values of ${\cal A}_{i}(q^2)$ are similar to
 $\varepsilon_{i}(q^2)$,  in the following numerical calculations,
 without loss of generality  we  concentrate on $\varepsilon_{i}(q^2)$.
 Moreover, we will not discuss the contributions from ${\cal O}_{3,4,7,8}$
 since they are very small.

\begin{figure}[tbp]
 \vspace{1cm} \centerline{ \psfig{figure=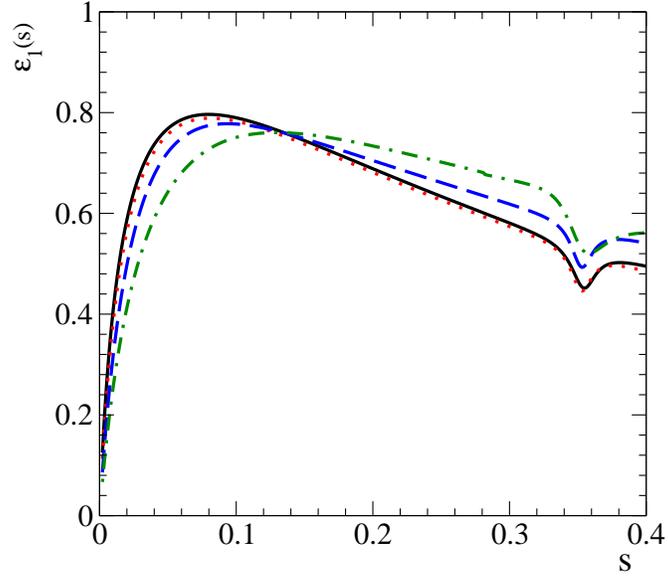,height=3.in } }
 \caption{Statistical significance $\varepsilon_1 (q^2)$
 of ${\cal O}_{1}$ as a
 function of $s=q^2/M^2_{B}$. The solid, dotted, dashed, and dash-dotted
curves
 stand for the results of the LCSR, QM, LFQM, and PQCD (I), respectively.
%while the dash-dotted one corresponds to that of the PQCD (I).
}
 \label{figo1}
 \end{figure}

\begin{figure}[tbp]
 \vspace{1.3cm} \centerline{ \psfig{figure=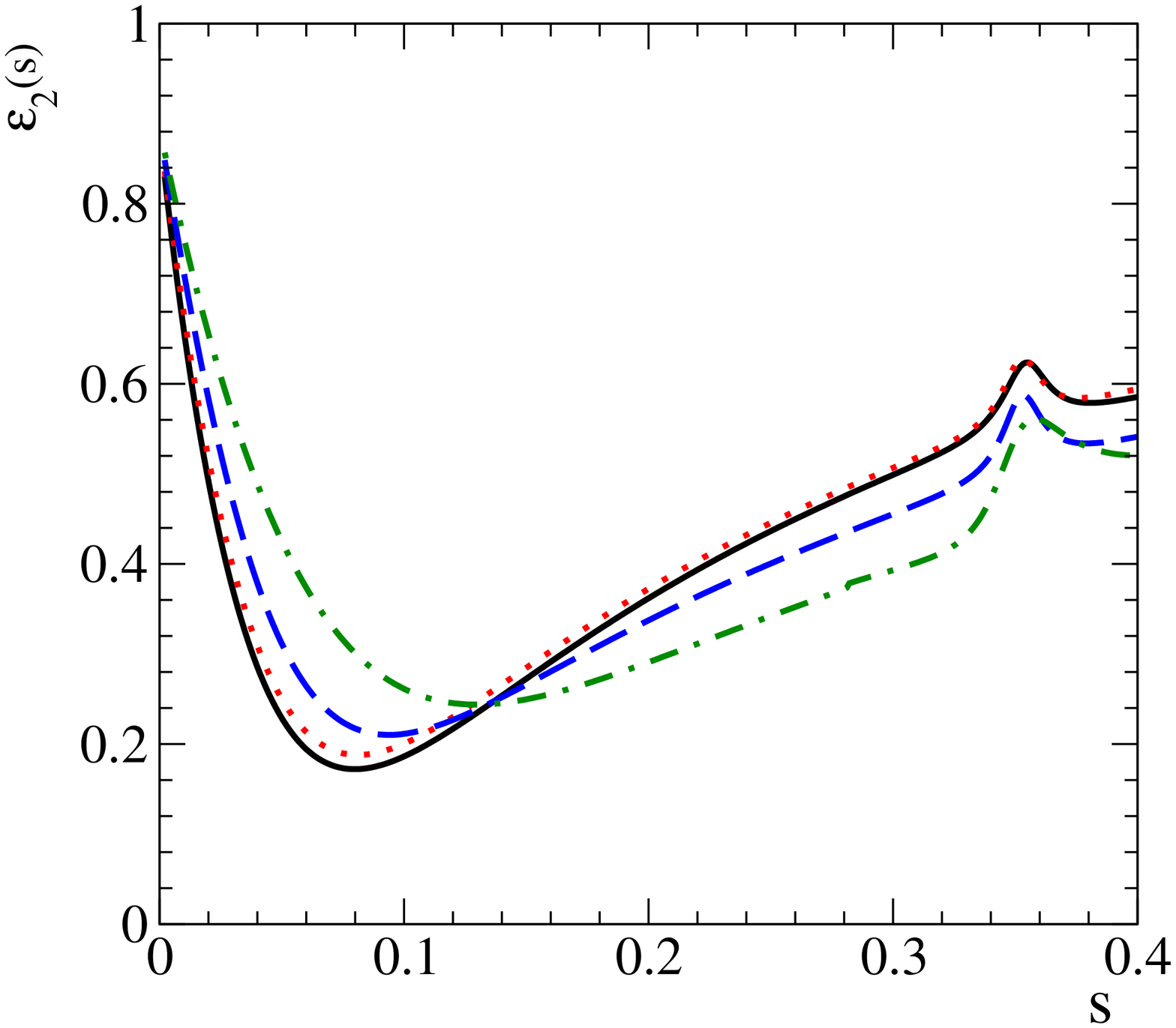,height=3.in } }
 \caption{Same as Figure \ref{figo1} but for $\varepsilon_2
 (q^2)$.}
 %The statistical significance of ${\cal O}_{2}$. Legend is
 %the same as figure \ref{figo1}. }
 \label{figo2}
 \end{figure}

 In Figures \ref{figo1}  and \ref{figo2}, we show the statistical
 significances of ${\cal O}_1$ and ${\cal O}_2$,
%which are also
 related to the longitudinal and transverse polarizations of $K^*$,
in various QCD approaches, respectively.
  From these figures, we
 see that the differences among different QCD approaches are
 insignificant, $i.e.$, they are not sensitive to hadronic effects
 so that they can be used as good candidates to test the SM as well as
 search for new physics.

\begin{figure}[tbp]
 \vspace{1cm} \centerline{ \psfig{figure=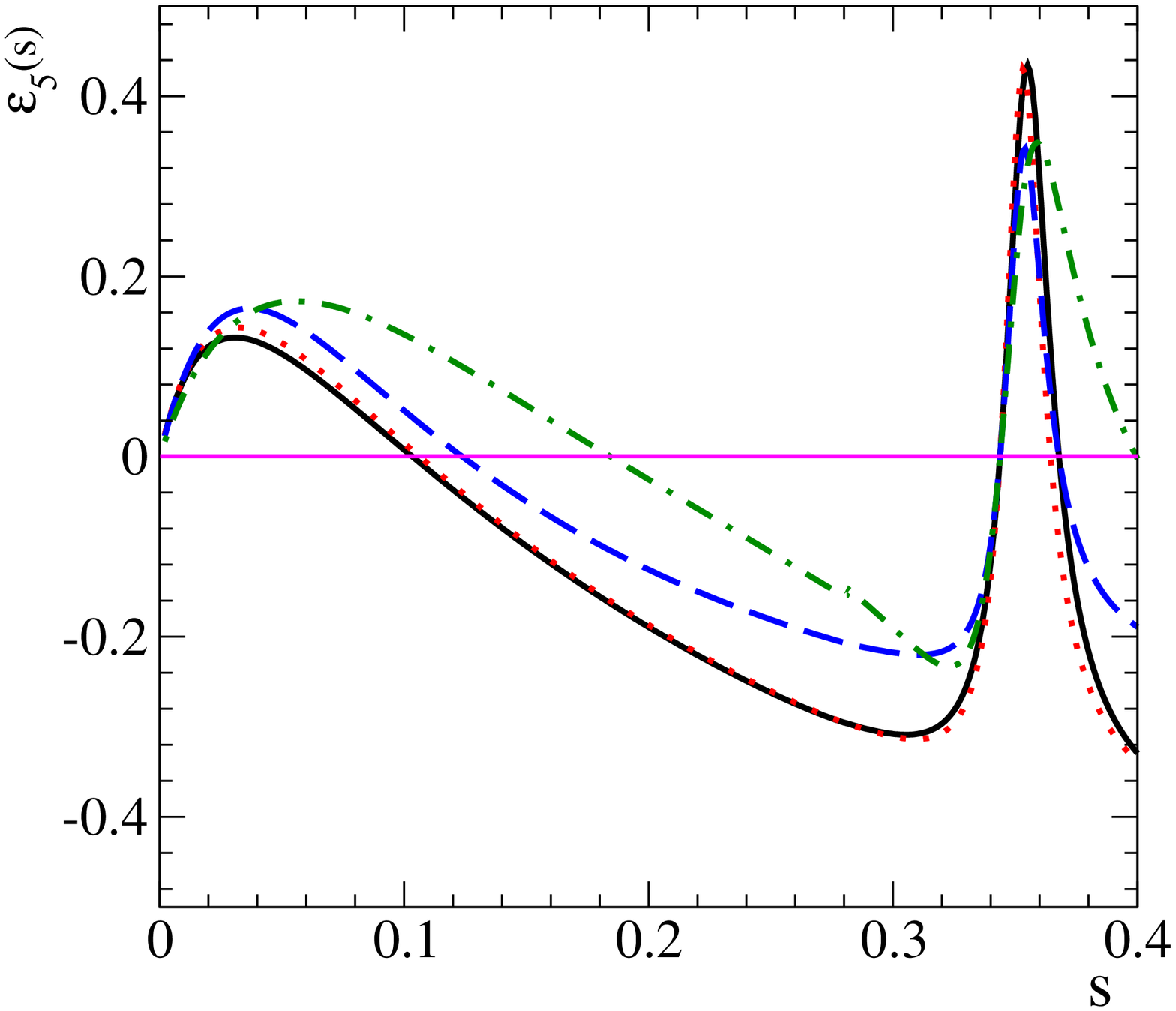,height=3.in } }
 \caption{Same as Figure \ref{figo1} but for $\varepsilon_5 (q^2)$.}
 %The statistical significance of ${\cal O}_{5}$. Legend is
 %the same as figure \ref{figo1}. }
 \label{figo5}
 \end{figure}

\begin{figure}[tbp]
 \vspace{1cm} \centerline{ \psfig{figure=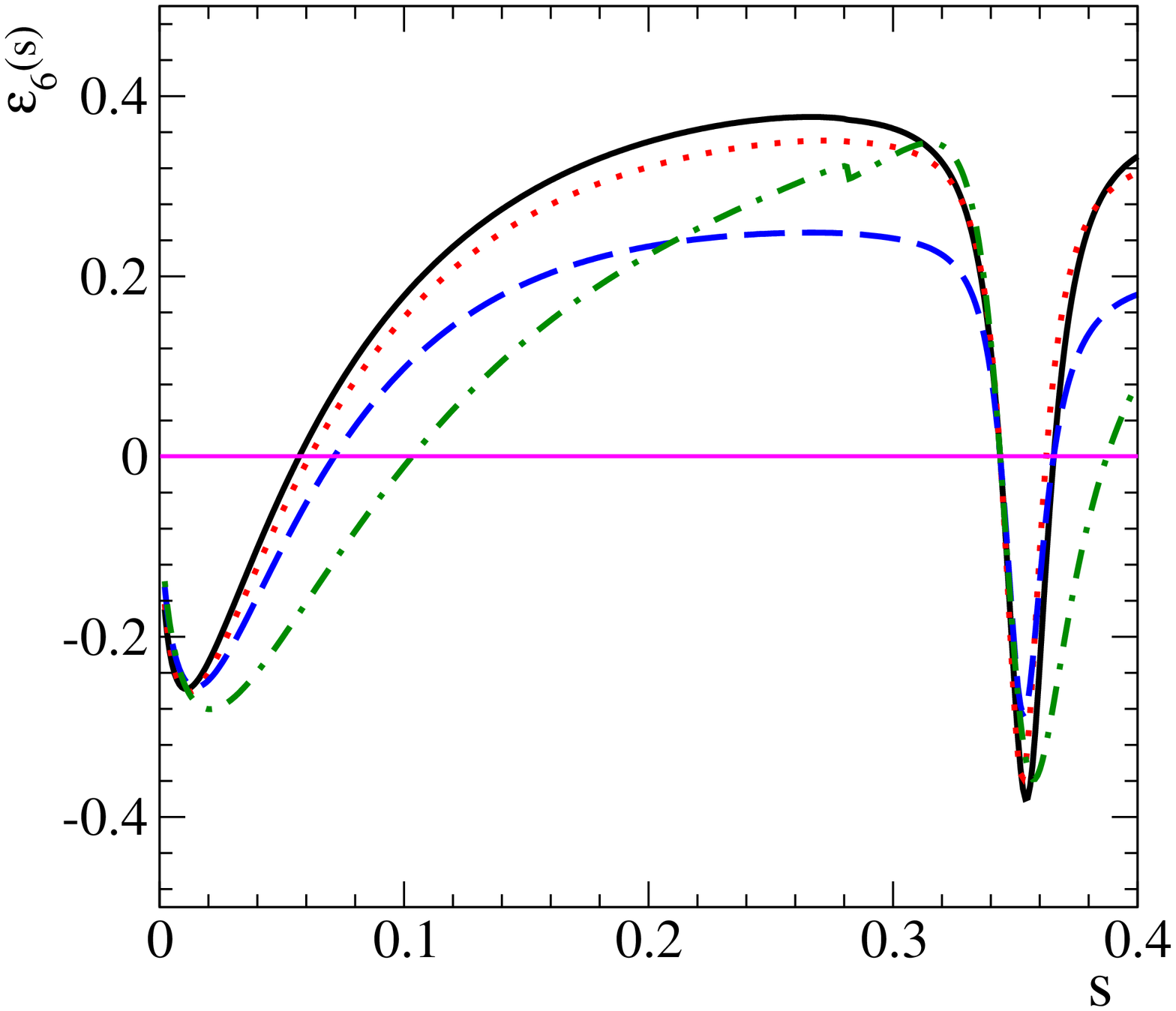,height=3.in } }
 \caption{Same as Figure \ref{figo1} but for $\varepsilon_6 (q^2)$.}
 %The statistical significance of ${\cal O}_{6}$. Legend is
 %the same as figure \ref{figo1}. }
 \label{figo6}
 \end{figure}

 In Figures  \ref{figo5} and \ref{figo6}, we display
 $\varepsilon_{5}(s)$ and $\varepsilon_{6}(s)$ as functions
 of $s=q^2/M_B^2$, which correspond to the angular distributions of
 $\cos\theta_{\ell}$ and $\sin2\theta_{K}$ in  Eq. (\ref{difangle})
 %and depend on the mixtures of the longitudinal (${\cal M}^{0}_{1(2)}$)
 %and transverse (${\cal M}^{\pm}_{2(1)}$)
 and depend both on $ReC^{eff}_9C^{*}_{10}$ and
 $ReC_{7}C^{*}_{10}$,
 respectively.
 %Their numerical results under different QCD
 %approaches are shown in Figures \ref{figo5} and \ref{figo6}.
  From the figures, we see that the zero points of
 $\varepsilon_{5}$ and $\varepsilon_{6}$ in the PQCD are quite
 different from others. Hence, by measuring these distributions,
especially those zero points,
  we can distinguish the PQCD results from other QCD models.

\begin{figure}[tbp]
 \vspace{1cm} \centerline{ \psfig{figure=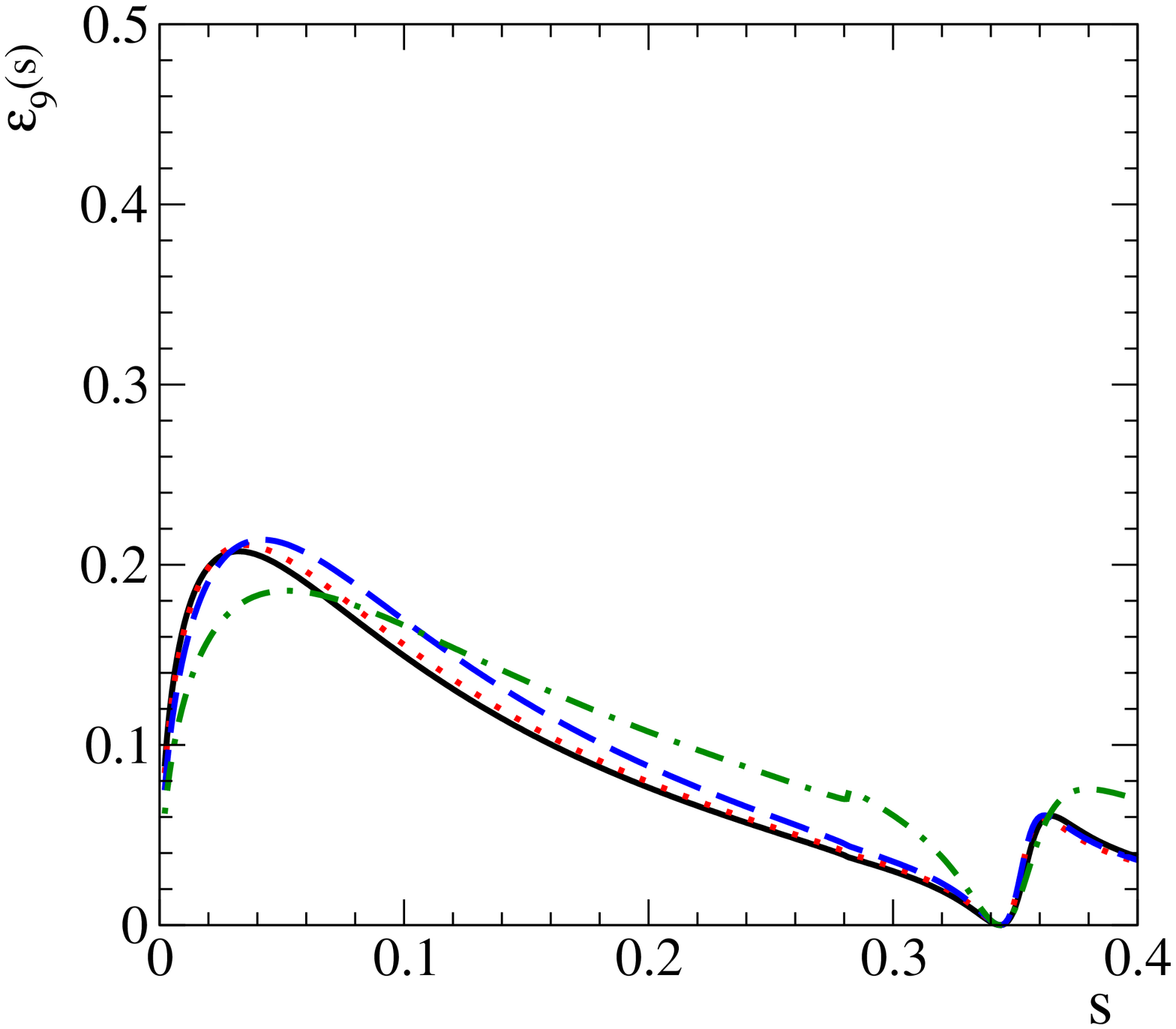,height=3.in } }
 \caption{Same as Figure \ref{figo1} but for $\varepsilon_9 (q^2)$
 and with $ImC_7=0.25$.}
 %The statistical significance of ${\cal O}_{9}$ with
 %$ImC_7=0.25$. Legend is the same as figure \ref{figo1}.
 %}
 \label{figo91}
 \end{figure}

\begin{figure}[tbp]
 \vspace{1cm} \centerline{ \psfig{figure=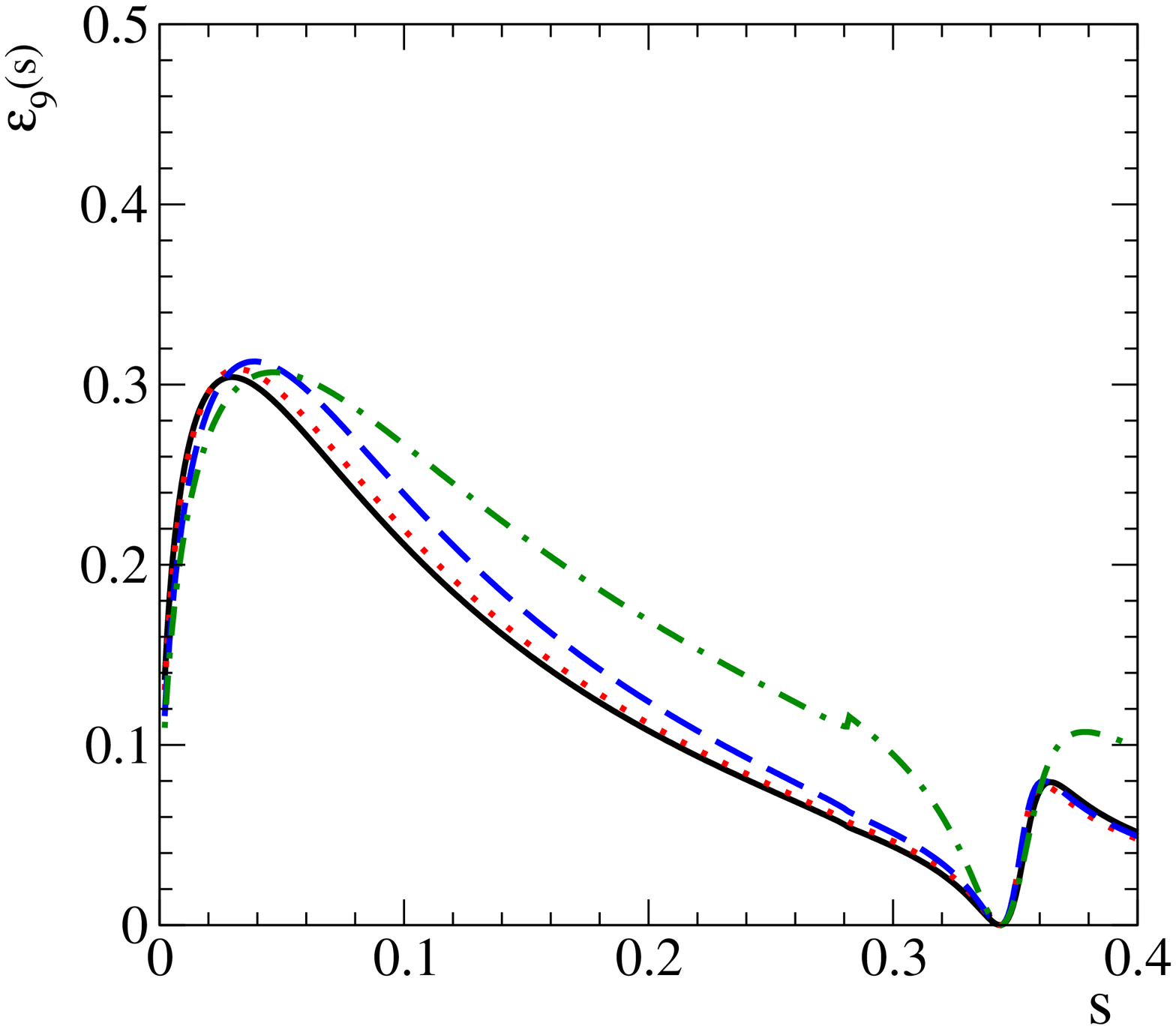,height=3.in } }
 \caption{Same as Figure \ref{figo1} but for $\varepsilon_9 (q^2)$
 and with $ImC_7=0.25$ and $ImC_{10}=-2.0$.}
 %\caption{The statistical significance of ${\cal O}_{9}$ with
 %$ImC_7=0.25$ and $ImC_{10}=-2.0$. Legend is the same as figure
 %\ref{figo1}. }
 \label{figo92}
 \end{figure}

 To show T violating effects, we concentrate on the T-odd operator of
${\cal O}_9$
% \cite{CG-EJP},
and consider new CP violating
sources beyond the CKM.
In the SM, the contribution to $\varepsilon_9$
is less than $O(1\%)$.
As illustrations, in Figures \ref{figo91} and \ref{figo92},
  we present our results by taking
 (i) $Im C_7=0.25$ and (ii) $Im C_7=0.25$ and $Im C_{10}=-0.20$
 with the others being the same as those in the SM, respectively.
 %in Figure \ref{figo91} and $Im C_7=0.25$ and $Im C_{10}=-0.20$
 %in Figure \ref{figo92}.
 One possible origin of having these
 imaginary parts is from SUSY
%\cite{geng-PRL}
  where there are many CP violating sources.
 It is interesting to see that the CP violating effect in Figure
 \ref{figo92} can be as
 large as $30\%$ in these models with new physics.
  %as shown in Figure \ref{figo92}.
  We emphasize that a measurement of such effect is a
clear indication of new physics as contrast with the decay rates
for which one could not distinguish the non-standard effect due to
the large uncertainties in various QCD models as shown in Figure
\ref{figdifbr}.
% Similarly,
Finally, we note that unlike $\varepsilon_{9}$, $\varepsilon_{7,8}$
receive contributions from the absorptive parts
in $C_9^{eff}(\mu)$ in the SM and they conserve CP.
% \cite{CG-EJP}.
On the other hand, they are much smaller than $\varepsilon_{9}$ in new
physics models
%contributions from models
such as the ones in (i) and (ii).
% comparing with $\varepsilon_{9}$.
Due to the uncertainty in the form of $C_9^{eff}(\mu)$ in Eq.
(\ref{effc9}), we shall not discuss them further.

 \section{Conclusions}
 We have studied the exclusive decays of $B\to K^{*}\ell^+ \ell^-$ within
 the framework of the PQCD. We have obtained the form factors for the
 $B\to K^{*}$ transition in the large recoil region, where the PQCD
 for heavy $B$ meson decays is reliable. We have found that the form
 factors at $q^2=0$ are consistent with those from most
 of the other QCD models, in particular, the
 LEET combined with the HQET and the experimental data on  $B\rightarrow
 K^* \gamma$. We have related  the angle distributions in
  the decay chains of $B\rightarrow K^* (K\pi) \ell^+
 \ell^-$ with 9
  physical observables due to  the different
 helicity combinations in $B\rightarrow K^* \ell^+ \ell^-$.
 In particular, we have shown that the T-odd observable can be used
 to test the SM and search for new physics, which is $<O(1\%)$ and up
 to  $O(10\%)$, respectively. Finally, we remark that to measure
 such an $10\%$ CP violating effect experimentally at the $1\,\sigma$
 level, for example, in $B\to K\pi \mu^+\mu^-$ one need at least
 $7.7\times 10^7\ B\bar{B}$, which is accessible in the current
 B factories at KEK and SLAC.\\

 \noindent {\bf Acknowledgments}

 We would like to thank K. Hagiwara, W.S. Hou, Y.Y. Keum, and H.N. Li
 for useful discussions. This work was supported in part by
 the National Science Council of the Republic of China under
 Contract Nos. NSC-90-2112-M-001-069 and NSC-90-2112-M-007-040 and
 the National Center for Theoretical Science.

\baselineskip=0.7 cm
 %%\begin{references}

% \begin{references}

%% \end{references}

 \newpage

 %%%%%%%%%%%%%%%%%%%%%%%%%%%%%%%%%%%%%%%%%%%%%%%%%%%%%%%%%%%%%
 %%%%%%%%%%%%%%%%%%%%%%%%%%%%%%%%%%%%%%%%%%%%%%%%%%%%%%%%%%%%%
 %%%%%%%%%%%%%%%%%%%%%%%%%%%%%%%%%%%%%%%%%%%%%%%%%%%%%%%%%%%%%
 %%%%%%%%%%%%%%%%%%%%%%%%%%%%%%%%%%%%%%%%%%%%%%%%%%%%%%%%%%%%%
 %%%%%%%%%%%%%%%%%%%%%%%%%%%%%%%%%%%%%%%%%%%%%%%%%%%%%%%%%%%%%
 %%%%%%%%%%%%%%%%%%%%%%%%%%%%%%%%%%%%%%%%%%%%%%%%%%%%%%%%%%%%%
 %%%%%%%%%%%%%%%%%%%%%%%%%%%%%%%%%%%%%%%%%%%%%%%%%%%%%%%%%%%%%
 %%%%%%%%%%%%%%%%%%%%%%%%%%%%%%%%%%%%%%%%%%%%%%%%%%%%%%%%%%%%%
 %%%%%%%%%%%%%%%%%%%%%%%%%%%%%%%%%%%%%%%%%%%%%%%%%%%%%%%%%%%%%
 %%%%%%%%%%%%%%%%%%%%%%%%%%%%%%%%%%%%%%%%%%%%%%%%%%%%%%%%%%%%%
 %%%%%%%%%%%%%%%%%%%%%%%%%%%%%%%%%%%%%%%%%%%%%%%%%%%%%%%%%%%%%
 %%%%%%%%%%%%%%%%%%%%%%%%%%%%%%%%%%%%%%%%%%%%%%%%%%%%%%%%%%%%%
 %%%%%%%%%%%%%%%%%%%%%%%%%%%%%%%%%%%%%%%%%%%%%%%%%%%%%%%%%%%%%
 %%%%%%%%%%%%%%%%%%%%%%%%%%%%%%%%%%%%%%%%%%%%%%%%%%%%%%%%%%%%%
 %%%%%%%%%%%%%%%%%%%%%%%%%%%%%%%%%%%%%%%%%%%%%%%%%%%%%%%%%%%%%
 %%%%%%%%%%%%%%%%%%%%%%%%%%%%%%%%%%%%%%%%%%%%%%%%%%%%%%%%%%%%%
%%%%%%%%%%%%%%%%%%%%%%%%%%%%%%%%%%%%%%%%%%%%%%%%%%%%%%%%%%%%
 \end{document}